\documentstyle[aps,prl,floats,graphicx,amssymb]{revtex}

\renewcommand{\textrm}{\rm} 

\newcommand{\ul}{\underline}
\newcommand{\td}{\tilde}

\newcommand{\Tr}{\mathop{\mathrm{Tr}}}

\newcommand{\dyadic}[1]{\stackrel{_\leftrightarrow}{#1}}

\newcommand{\vecb}[1]{{\bf{#1}}}
\newcommand{\upd}{{\mathrm{d}}}
\newcommand{\nvec}{\hat{\vecb n}}
\newcommand{\svec}{\hat{\vecb s}}
\newcommand{\kvec}{{\hat{\vecb k}}}
\newcommand{\xvec}{\hat{\vecb x}}
\newcommand{\yvec}{\hat{\vecb y}}
\newcommand{\zvec}{\hat{\vecb z}}

\newcommand{\rvec}{\vecb r}
\newcommand{\gvec}{\vecb g}
\newcommand{\fvec}{\vecb f}
\newcommand{\avec}{\vecb a}
\newcommand{\bvec}{\vecb b}
\newcommand{\cvec}{\vecb c}
\newcommand{\Cvec}{\vecb C}

\newcommand{\dvec}{\vecb d}

\newcommand{\nuvec}{\bbox{\nu}}
\newcommand{\Deltavec}{\bbox{\Delta}}
\newcommand{\sigmavec}{\bbox{\sigma}}
\newcommand{\cvecim}{\mathop{\textrm{Im}}\cvec}
\newcommand{\vF}{v_{\mathrm{F}}}
\newcommand{\NF}{N(0)}
\newcommand{\kB}{k_{\mathrm{B}}}

\newcommand{\iu}{{\mathrm{i}}}
\newcommand{\me}{\epsilon_m}
\newcommand{\nam}{\breve} 
\newcommand{\Sgn}{\mathop{\textrm{Sgn}}} 
\newcommand{\Det}{\mathop{\textrm{Det}}} 
\newcommand{\nc}{s}

\newcommand{\re}{\mathop{\textrm{Re}}}

\newcommand{\mene}{\epsilon_m}
\newcommand{\jfront}{2m_3\vF\NF\pi\kB T}
\newcommand{\efront}{\hbar\vF\NF\pi\kB T}

\begin{document}
\draft
\twocolumn[\hsize\textwidth\columnwidth\hsize\csname 
@twocolumnfalse\endcsname

\title{Pinhole calculations of the Josephson effect in $^3$He-B}

\author{J. K. Viljas$^{1}$ and E. V. Thuneberg$^{1,2}$}
\address{$^1$Low Temperature Laboratory, Helsinki University of
Technology, P.O.Box 2200, FIN-02015 HUT, Finland } 
\address{$^2$Department of Physical Sciences,
P.O.Box 3000, FIN-90014 University of Oulu, Finland } 

\date{\today}

\maketitle

\begin{abstract}
We study theoretically the dc Josephson effect between two volumes of
superfluid
$^3$He-B. We first discuss how the calculation of the
current-phase relationships is divided into a mesoscopic and a
macroscopic problem. We then analyze mass and spin currents and the
symmetry of weak links. In quantitative calculations the weak link is
assumed to be a {\it pinhole}, whose size is small in comparison to the
coherence length. We derive a quasiclassical expression for the
coupling energy of a pinhole, allowing also for scattering in the hole.
Using a selfconsistent order parameter near a wall, we calculate the
current-phase relationships in several cases. In the {\it isotextural}
case, the current-phase relations are plotted assuming a constant
spin-orbit texture. In the opposite {\it anisotextural} case the
texture changes as a function of the phase difference. For that we have
to consider the stiffness of the macroscopic texture, and we
also calculate some surface interaction  parameters. We analyze the
experiments by Marchenkov \emph{et al}. We find that the
observed $\pi$ states and bistability hardly can be explained with the
isotextural pinhole model, but a good quantitative agreement is
achieved with the anisotextural model. 
\end{abstract}

\bigskip
] 

\section{Introduction} \label{s.intro}

The Josephson effects in superconducting weak links have been actively
studied and applied since 1960's. By analogy, similar effects  also
exist between two volumes of superfluid connected by a weak link. 
There has been recent progress in observing the Josephson effect in
superfluid $^4$He
\cite{dc4he}. In this paper we study superfluid $^3$He,  where the
Josephson effect was experimentally confirmed over ten years ago by
Avenel and Varoquaux
\cite{original2,original}. However, more recent experiments in Berkeley
have raised new questions 
\cite{berkeley,bistability} which we are going to address here. 
 
Avenel and Varoquaux \cite{original2,original} used a single narrow
slit as the weak link in $^3$He. They found
current-phase relations
$J(\phi)$ that are very similar to those seen
for tunneling junctions or microbridges in $s$-wave superconductors. In
these systems the relations are generally close to a sine function,
$J(\phi)=J_{\rm c}\sin\phi$, or slightly tilted from this form. They are
characterized by a single maximum supercurrent $J_{\rm c}$ and a negative
derivative at $\phi=\pi$, $J'(\pi)<0$.  The experiments
at Berkeley used a $65\times 65$ array of
small apertures in $^3$He-B \cite{berkeley,bistability}. At high
temperatures
$J(\phi)$ was found to be sinusoidal. At lower temperatures a ``$\pi$
state'' developed, where the derivative
is positive at $\phi\approx \pi$: $J'(\pi)>0$. In addition, the weak
link could be found in two distinct states with different 
current-phase relations. One of the ``bistable'' states had consistently
higher critical current (H state) than the other (L state). Preliminary
results of $\pi$ states and multistability in a single narrow
slit have also been reported
\cite{singleaperture}.

Several theories have been put forward to explain these findings
\cite{hatakenaka,bose,spoil,yippi,viljas}. It
has been suggested that the reduction of the Josephson coupling due to
finite number of particles can lead to a $\pi$ state in trapped atomic
gases \cite{SFGS97}. This suggestion has been extended to $^3$He
\cite{hatakenaka,bose}. In the present paper we do calculations
with the quasiclassical theory, which is an exact expansion in
$T_{\rm c}/T_{\rm F}$ (the superfluid transition temperature over the Fermi
temperature), and find no sign of such a mechanism in the leading order.
Therefore we consider it unlikely that this mechanism could
quantitatively explain the $\pi$ state observed in 
$^3$He. What look more
promising are theories based on the $3\times 3$
matrix structure of the order parameter in $^3$He
\cite{yippi,viljas,MT86,thu88}. Unusual current-phase relations in
$^3$He were first calculated by Monien and Tewordt \cite{MT86}.
Their calculation used a very simplified one-dimensional
Ginzburg-Landau model, and the physical relevance of their intermediate
branches around $\phi\approx\pi$ remains controversial. The first
unambiguous evidence of a new branch in $J(\phi)$ came from
2-dimensional Ginzburg-Landau calculations
\cite{thu88}. Besides the usual case of parallel $\hat{\bf n}$
vectors on the two sides of the junction, this calculation
considered also antiparallel $\hat{\bf n}$
vectors, and the unusual $J(\phi)$ was found only in the latter case.
The new branch in $J(\phi)$ did not yet qualify as a $\pi$ state,
however, because $J'(\pi)$ was found negative at the parameter values
studied in Ref.\ \cite{thu88}. More extensive Ginzburg-Landau studies
in Ref.\ \cite{viljas} found that  a proper $\pi$ state ($J'(\pi)>0$)
occurs in the case of {\it parallel}
$\hat{\bf n}$'s through spontaneous symmetry breaking in a sufficiently
large aperture.

The Ginzburg-Landau calculations are applicable to relatively large
apertures. A tractable opposite limit is a very small aperture, a
pinhole. The pinhole model was first studied by Kulik and  Omel'yanchuk
for an \emph{s}-wave superconductor \cite{kulik}. At low temperatures
$J(\phi)$ has considerable deviation from the sine form, but there is no
$\pi$ state. Kurkij\"arvi considered the same problem in
$^3$He \cite{kurkijarvi}. In $^3$He-B the order parameter is always
modified near surfaces. Neglecting this complication, Kurkij\"arvi
found that $J(\phi)$ for parallel $\hat{\bf n}$'s is exactly the same
as for \emph{s}-wave superconductors. Yip generalized Kurkij\"arvi's
calculation to other orientations of the $\hat{\bf n}$ vectors
\cite{yippi}. He found a
$\pi$ state for antiparallel $\hat{\bf n}$'s, as well as for some more
complicated configurations, which can occur in a magnetic field
$H\gtrsim 1$ mT. We call this mechanism of the
$\pi$ state {\it isotextural} because the texture [the field
$\hat{\bf n}({\bf r})$] is kept constant while calculating $J(\phi)$.

The discussion above concerned a single aperture. There exist three
different suggestions as to how a $\pi$ state can appear in an array of
apertures. Avenel \emph{et al.}\ assumed that if the individual apertures
have a hysteretic $J(\phi)$, approximately half of the apertures could
be on a different branch than the others. 
The net effect would be the formation of a $\pi$ state \cite{spoil}. We consider
this explanation unlikely because apparently the apertures in  
Ref.\ \cite{bistability} are not hysteretic, and also because it is difficult
to understand why exactly half of the apertures could behave
differently. The second alternative is that the $\pi$ state appears
trivially in a coherent array if an isotextural $\pi$ state appears
in each of the apertures independently. The
third alternative is an {\it anisotextural}
$\pi$ state, where the $\hat{\bf n}$ texture changes as a function of
$\phi$
\cite{viljas}. This mechanism can lead to a $\pi$ state even in the
case when the isotextural $J(\phi)$ is sinusoidal. 

The purpose of this paper is to study $J(\phi)$ in $^3$He-B as
completely as possible using the pinhole model. Section \ref{s.symm}
starts with a division of the problem into mesoscopic and
macroscopic, and using symmetry arguments derives general expressions
for the Josephson coupling.  The mesoscopic problem is discussed in
the following four sections. Sec.\ \ref{s.quasi}
introduces the quasiclassical theory and the assumptions relevant for
our calculation. The pinhole model is defined in Sec.\ \ref{s.pin} and
a general pinhole energy functional is derived Sec.\ \ref{s.ene}.
The propagators are calculated in Sec.\ \ref{s.phcalc}, and the
Josephson energy and the currents are evaluated. This corrects the
calculations by Kurkij\"arvi
\cite{kurkijarvi} and by Yip \cite{yippi} by using a selfconsistently
calculated order parameter. We consider both diffuse and specular
reflection of quasiparticles at the wall. We also discuss the case
where scattering is present within the pinhole. 

The rest of the paper is
devoted to the macroscopic problem. In Sec.\
\ref{s.surf} we discuss the interactions that are important on the
macroscopic scale. We estimate the length scales and evaluate some
surface-interaction parameters that have not been calculated before. In
Sec.\ \ref{s.pinres} we plot isotextural
current-phase relations in different
situations. The
anisotextural Josephson effect is discussed in Sec.\ \ref{s.synerg}.
The effect is first demonstrated with a simple model. Then we
estimate the textural rigidity, and calculate current-phase
relationships for an array of
pinholes. Section
\ref{s.est} is devoted to the analysis of different $\pi$ state models
in the Berkeley experiment
\cite{bistability}. Section \ref{s.conc} finishes
with some conclusions.

\section{Symmetry considerations} \label{s.symm}

The Cooper pairs in superfluid $^3$He are in a relative $p$ wave state 
and form a spin triplet. This state of affairs is reflected by the 
$3\times 3$ tensor character of the order parameter $A_{\mu i}$.  The
first index in $A_{\mu i}$ (greek letter) refers to the spin states and
the second index (latin letter) to the orbital states. In the bulk of
$^3$He-B the order parameter has the form \cite{vollhardt}
\begin{equation} \label{e.opa}
A_{\mu i}=R_{\mu i}\Delta\exp(\iu\phi).
\end{equation}
Here $\Delta$ is a real constant and $R_{\mu i}$ is a rotation matrix
satisfying
$R_{\mu i}R_{\mu j}=\delta_{ij}$. (We shall consistently assume a 
summation over $x$, $y$, and $z$ for repeated index variables
$\mu$, $\nu$, $i$, $j$, $\alpha$, $\beta$, $\gamma$, etc.) The
rotation-matrix structure and $\Delta$ are fixed on the scale of the
superfluid condensation energy. On this energy scale, the state
(\ref{e.opa}) remains degenerate with respect to the phase
$\phi$ and different rotations $R_{\mu i}$, which are parametrized with
an axis
$\nvec$ and an angle $\theta$. The degeneracy with respect to
$\phi$ and $R_{\mu i}$ (or equivalently $\phi$, $\nvec$, and
$\theta$) is partly lifted by weaker interactions, which are discussed
in detail in Sec.\ \ref{s.surf}. 

\begin{figure}[!tb]
\begin{center}
\includegraphics[width=0.7\linewidth]{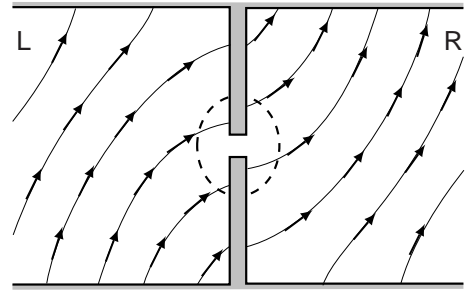}
\caption{A weak link between two bulk superfluids, $L$ and $R$.  The
arrows denote the $\nvec$ vector of $^3$He-B. The dashed line
divides the liquid into a mesoscopic region (inside) and
a macroscopic region (outside).}
\label{f.junctionregions}
\end{center}
\end{figure}

Imagine now two bulk volumes of $^3$He-B connected by a weak link, see
Fig.\ \ref{f.junctionregions}. 
We consider the system to consist of a mesoscopic region at the
junction and a macroscopic region outside. There can be considerable
variation of the soft degrees of freedom in the macroscopic region, as
illustrated for the $\nvec$ vector in Fig.\ \ref{f.junctionregions}.
The mesoscopic region is chosen sufficiently small so that (i) 
both $\phi$ and
$R_{\mu i}$ are effectively constants at the macroscopic-mesoscopic
border, and (ii) the weak interactions affecting $\phi$ and
$R_{\mu i}$ can be neglected in the mesoscopic region. The size of the
mesoscopic region is limited from below by the condition that the bulk
form of the order parameter (\ref{e.opa}) has to be valid in the
macroscopic region. Thus the size of the mesoscopic region has to
be large in comparison to both the superfluid coherence length
$\xi_0\approx 10$ nm and the size of the aperture. The mesoscopic
region can also be chosen to cover several apertures. 

The rotation matrices and the phases on the left ($L$) and right ($R$)
sides are generally different. Their values at the
macroscopic-mesoscopic border are denoted by 
$R_{\mu i}^{L,R}$ and $\phi^{L,R}$.  The most general form for the
Josephson coupling energy associated with the weak link is given by
\begin{equation} \label{e.form1}
 F_{\rm J}=F_{\rm J}(\phi^L,\phi^R,R_{\mu i}^L,R_{\nu j}^R).
\end{equation} However, due to global phase
invariance there can really be only a dependence on the phase
difference $\phi\equiv\phi^R-\phi^L$. In addition, if we assume the
intervening wall to be magnetically inactive, then we should have
invariance with respect to global spin rotations as well. This means
that the energy can only depend on the  product of the rotation
matrices, or the quantities 
$\psi_{i j}=R_{\mu i}^LR_{\mu j}^R$, so that
\begin{equation} \label{e.form2}
 F_{\rm J}=F_{\rm J}(\phi,R_{\mu i}^LR_{\mu j}^R).
\end{equation}
Using the functional $F_{\rm J}$ (\ref{e.form2}) one can now calculate
two conserved currents. Firstly, the mass current through the
aperture is given by
\begin{equation} \label{e.enederiv}
J=\frac{2m_3}{\hbar}\frac{\partial F_{\rm J}}{\partial\phi}.
\end{equation}
See, for example, Ref.\ \cite{anderson}.
Secondly, due to the different spin-orbit rotation matrices on the two
sides, there is also a spin current flowing through the aperture.
Likewise, this is obtained from $F_{\textrm{J}}$ by differentiation:
\begin{equation} \label{e.plausible}
J^{\textrm{spin}}_\gamma=\epsilon_{\alpha\beta\gamma}
R_{\alpha i}^LR_{\beta j}^R
\frac{\partial F_{\rm J}}{\partial(R_{\mu i}^LR_{\mu j}^R)}.
\end{equation}
This can be seen by replacing $\psi_{i j}$ in (\ref{e.form2}) by
$R_{\alpha i}^LR_{\beta j}^R R_{\alpha \beta}$ with the relative
rotation
$R_{\alpha\beta}=\delta_{\alpha\beta}+\epsilon_{\alpha\beta\gamma}
\delta\theta_{\gamma}$, and identifying
$\delta F_{\rm J}
=J_\gamma^{\textrm{spin}}\delta\theta_\gamma+O(\delta\theta_\gamma^2)$
\cite{cross}. 

The energy and the currents also satisfy some other symmetry relations.
For example, assuming invariance with respect to time reversal, we have
$F_{\rm J}(-\phi,\psi_{ij})=F_{\rm J}(\phi,\psi_{ij})$. As a consequence
$J(-\phi,\psi_{ij})=-J(\phi,\psi_{ij})$ and
$J^{\textrm{spin}}_\gamma(-\phi,\psi_{ij})=
J^{\textrm{spin}}_\gamma(\phi,\psi_{ij})$. Also, since the phase factor
$\exp(\iu\phi)$ is defined only modulo $2\pi$, we have $F_{\rm
J}(\phi+2\pi,\psi_{ij})=F_{\rm J}(\phi,\psi_{ij})$. Here we must keep
in mind that in a long junction $F_{\rm J}$ is not a single-valued
function. Analogously, there is periodicity with respect to the rotation
angle, which is automatically contained in the form of Eq.\ 
(\ref{e.form1}), since
$R_{\mu i}(\nvec,\theta+2\pi)=R_{\mu i}(\nvec,\theta)$. 

The functional form of $F_{\rm J}$ (\ref{e.form2}) can be restricted
further if we assume some additional symmetries. For example, if the
aperture is symmetric under a parity operation, we have 
$F_{\rm J}(\phi,\psi_{ji})=F_{\rm J}(\phi,\psi_{ij})$ and thus
$J_\gamma^{\textrm{spin}}(\phi,\psi_{ji})=
-J_\gamma^{\textrm{spin}}(\phi,\psi_{ij})$.
If the aperture has
full ``orthorhombic'' symmetry
$\frac{2}{m}\frac{2}{m}\frac{2}{m}$, then $F_{\rm J}$ can only depend
on the rotation matrices through 
$\psi_{xx}$, $\psi_{yy}$, and $\psi_{zz}$. From here on we fix the 
$z$ coordinate to be along the axis of the weak link. Now, if the twofold
rotation symmetry around $z$ is replaced by  a fourfold symmetry
($\frac{4}{m}\frac{2}{m}\frac{2}{m}$),  then an exchange of $\psi_{xx}$ 
and
$\psi_{yy}$ must not affect $F_{\rm J}$.  Finally, if the rotation symmetry
around $z$ is continuous ($\frac{\infty}{m}\frac{2}{m}$),
the dependence can only be
through $\psi_{zz}$ and the invariant combination
$\psi_{xx}+\psi_{yy}$, that is
\begin{equation} \label{e.form3}
F_{\rm J}=F_{\rm J}(\phi,R_{\mu x}^LR_{\mu x}^R+R_{\mu y}^LR_{\mu y}^R,
R_{\mu z}^LR_{\mu z}^R).
\end{equation}

Close to $T_{\rm c}$ the amplitude $\Delta$ of the order parameter
(\ref{e.opa}) approaches zero, so that we can expand $F_{\rm J}(A_{\mu
i}^L,A_{\nu j}^R)$ in powers of $\Delta$. In order to be consistent with
Eq.\ (\ref{e.form3}), the leading order term in the expansion must be
\begin{equation} \label{e.fj} F_{\rm J}=-[\alpha R_{\mu z}^LR_{\mu z}^R+
\beta(R_{\mu x}^LR_{\mu x}^R+R_{\mu y}^LR_{\mu y}^R)]\cos\phi,
\end{equation} 
 where $\alpha$ and $\beta$ are some real valued
phenomenological constants. In Ref.\ \cite{viljas} this was introduced
as the Josephson energy of the tunneling model, but as the derivation
above shows, it is more general.  The tunneling barrier for $^3$He was
first considered in Ref.\ \cite{degennes}.

In order to further determine the functional $F_{\rm J}$
(\ref{e.form2},\ref{e.form3},\ref{e.fj}) it is necessary to do a
calculation in the mesoscopic region. This is discussed in the
following sections \ref{s.quasi}-\ref{s.phcalc}. 

\section{Quasiclassical theory} \label{s.quasi}

We use the quasiclassical theory of Fermi liquids \cite{sr} 
to calculate $F_{\rm J}$
(\ref{e.form2}) for a pinhole aperture. Here we present the theory
only in the depth needed for the following.  

The central quantity is the
quasiclassical propagator $\nam g$.  In the
stationary case which we are considering this can be written as
$\nam g(\kvec,\rvec,\mene)$,  where $\rvec$ denotes spatial position,
$\kvec$ parametrizes a position on the Fermi surface, 
and $\mene=\pi\kB T(2m+1)$ are the
Matsubara energies.  The propagator is determined by the 
Eilenberger equation and
the normalization condition
\begin{eqnarray} [\iu\mene \nam{\tau}_3-\nam{\sigma},\nam{g}]+
\iu\hbar\vF\kvec\cdot\nabla_{\rvec} \nam{g} & = &  0  \label{e.eil} \\ 
\nam{g} \nam{g} & = & -\nam{1}.
\end{eqnarray} Equation (\ref{e.eil}) can be interpreted as 
describing transport
of quasiparticle wave packets which travel on classical trajectories
with the Fermi  velocity $\vecb v_{\textrm{F}}=\vF\kvec$. The
propagator $\nam g$ as well as the self-energy $\nam
\sigma(\kvec,\rvec)$ are $4\times4$ matrices, reflecting the spin and
particle-hole degrees of freedom of a quasiparticle; the matrices
$\nam \tau_i$ $(i=1,2,3)$ are the Pauli matrices in the particle-hole
space. 

The $2\times2$ spin blocks of $\nam g$ can be decomposed
into scalar and vector components as
\begin{equation} 
\nam{g}  = \left[ \begin{array}{cc} \label{e.gdecomp}
g+\gvec\cdot\ul{\sigmavec} &
(f+\fvec\cdot\ul{\sigmavec})\iu\ul{\sigma}_2\\
\iu\ul{\sigma}_2(\td f + \td\fvec\cdot\ul{\sigmavec}) & 
\td g - \ul{\sigma}_2 \td\gvec\cdot\ul{\sigmavec}\,\ul{\sigma}_2
\end{array} \right],
\end{equation}
where
$\ul\sigmavec=\xvec\ul\sigma_1+\yvec\ul\sigma_2+\zvec\ul\sigma_3$,  and
$\ul\sigma_i$ $(i=1,2,3)$ are the spin-space Pauli matrices. The
self-energy $\nam\sigma$ is written similarly
\begin{equation}
\nam \sigma  = \left [ \begin{array}{cc} \label{e.edecomp}
\nu+\nuvec\cdot\ul\sigmavec & \Deltavec\cdot\ul\sigmavec\iu\ul\sigma_2
\\
\iu\ul\sigma_2\Deltavec^*\cdot\ul\sigmavec & 
\td\nu - \ul\sigma_2 \td{\nuvec}\cdot\ul\sigmavec\,\ul\sigma_2
\end{array} \right]. 
\end{equation}
Here the off-diagonal terms contain the $p$-wave pairing interaction
in the form of the gap vector
$\Delta_\mu(\kvec,\rvec)=A_{\mu i}(\rvec)\hat k_i$. 
This is determined by the self-consistency equation
\begin{eqnarray} \label{e.selfconsist}
\pi \kB T & & \sum_m \left [
\frac{\Deltavec}{|\epsilon_m|}-3\int\frac{\upd\Omega_{\kvec'}}{4\pi}
\fvec(\kvec',\rvec,\epsilon_m)(\kvec'\cdot\kvec) \right] \nonumber\\
&&+\Deltavec\ln\frac{T}{T_{\rm c}} = \vecb 0,
\end{eqnarray}
where $T_{\rm c}$ is the superfluid transition temperature.
This form is valid in the weak coupling
approximation, where the quasiparticle-quasiparticle scattering is
neglected. The diagonal components $\nu,\td\nu$ and 
$\nuvec,\td{\nuvec}$ are the
``molecular field'' self-energies arising from redistributions of
quasiparticles. The scalar parameters $\nu$ and $\td\nu$ arise in 
response to mass currents, and they turn out to be negligible as will
be argued in Sec.\ \ref{s.pin}. The (real valued) 
vector parameters $\nuvec$ and $\td{\nuvec}$ describe the response 
to a magnetic field or spin
currents. As discussed above, the magnetic field can be neglected in the
mesoscopic region. In contrast, there are always rather strong spin
currents flowing along surfaces in 
$^3$He-B \cite{zhang}. These have to be taken into account with 
the self-consistency relation ($\td{\nuvec}=-\nuvec$)
\begin{equation} \label{e.nuself}
\nuvec(\kvec,\rvec)=\pi\kB T\sum_m\int\frac{\upd\Omega_{\kvec'}}{4\pi}
A^a(\kvec\cdot\kvec')\gvec(\kvec',\rvec,\epsilon_m).
\end{equation} 
Here $A^a(x)=\sum_{l=0}^\infty F^a_l[1+F^a_l/(2l+1)]^{-1}P_l(x)$,
$P_l$ are the Legendre polynomials, and all terms with even $l$ drop
out due to symmetries.

For our purposes the most important physical quantities 
to be evaluated from $\nam g$ are the mass supercurrent
\begin{equation} \label{e.genecurrent}
\vecb{j}(\rvec)=\jfront \sum_m \int
\frac{\upd\Omega_\kvec}{4\pi}\kvec g(\kvec,\rvec,\epsilon_m)
\end{equation}
and the spin supercurrent
\begin{equation} \label{e.spincurrent}
\vecb j^{\textrm{spin}}_{\gamma} (\rvec)=
\efront\sum_m\int\frac{\upd\Omega_\kvec}{4\pi}\kvec
g_\gamma(\kvec,\rvec,\epsilon_m).
\end{equation}
 Here
$N(0)=m^*{}^2\vF/(2\pi^2\hbar^3)$ is the one-spin normal-state density
of states at the Fermi surface, where $m^*=m_3(1+F_1^s/3)$ is the
effective quasiparticle mass, $m_3$ being the mass of a bare $^3$He
atom. The superfluid coherence length is defined by
$\xi_0=\hbar\vF/(2\pi\kB T_{\rm c})$. For $F_1^s$, $\vF$, $T_{\rm c}$ and other
pressure dependent  quantities we use the vapor pressure values
whenever needed. In Eq.\ (\ref{e.nuself}) we assume $F_l^a=0$ for all
odd $l\geq 3$. Since the parameter $F^a_1$ is not well known, we usually
set it to zero also, but values in the range  $-1...0$ have been tested.

\section{The pinhole problem} \label{s.pin}

\begin{figure}[!bt]
\begin{center}
\includegraphics[width=0.9\linewidth]{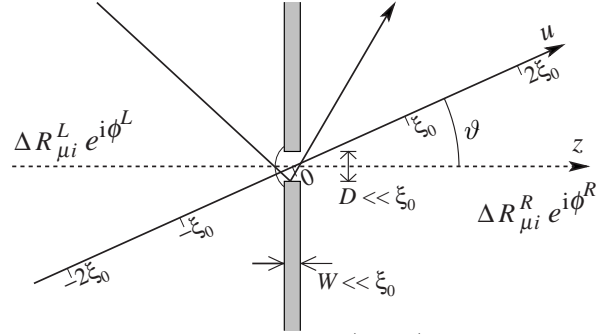}
\caption{
The mesoscopic region (Fig.\ \protect\ref{f.junctionregions}) for a
pinhole. Two quasiparticle trajectories are shown. The coordinate
$u$ is plotted along the straight transmitting trajectory. The arch on
the left hand side of the pinhole denotes an imaginary surface that
is used to close the pinhole (Sec.\ \protect\ref{s.ene}).  The diameter
$D$ of the pinhole and the thickness of the wall
$W$ are assumed small in comparison to $\xi_0$. }
\label{f.trajectory}
\end{center}
\end{figure}

Consider the case of a single pinhole in a thin wall separating
two volumes of $^3$He-B --- the situation of Fig.\ \ref{f.trajectory}.
The hole can be thought of as a pinhole (i.e. ``very small'') and still
be treated quasiclassically, if
its dimensions (diameter $D$ and wall thickness $W$) satisfy 
$\xi_0\gg D,W \gg \lambda_{\textrm{F}}$, where $\lambda_{\textrm{F}}$
is the Fermi wavelength. In addition $^3$He is in the pure limit,
where the mean free path of quasiparticles $l\gg\xi_0$. Usually one
further assumes that $W\ll D$ so that scattering in the aperture
itself can be neglected.  We allow for a finite $W/D$ and thus consider
also deflected trajectories of the form shown in
Fig.\ \ref{f.trajectory}. The pinhole limit was first considered within
the quasiclassical theory by Kulik and Omel'yanchuk in the case of 
$s$-wave superconducting  microbridges \cite{kulik}.  Several previous
calculations for the spin-triplet case of 
$^3$He also exist \cite{yippi,viljas,kurkijarvi,kop86,tks,rom,verditzK}. 
For another type of quasiclassical Josephson model, see Ref. \cite{rainerlee}.

What makes the pinhole case attractive is that {\it no} self-consistent
calculation of the order parameter in the aperture is needed. More
precisely, the leading term in the coupling energy $F_{\rm J}$ is on
the order of the superfluid condensation energy in the volume
$D^2\xi_0$. This effective volume is large compared to the volume ($\sim
D^3$ or $WD^2$) of the pinhole. The leading term in $F_{\rm J}$ 
(and thus also the corresponding terms in the currents) can
be computed using $\nam \sigma$ (\ref{e.edecomp}) that is calculated 
for a planar wall {\it
without} the pinhole. There {\it is} a correction $\propto D^2$ to
$\nam
\sigma$, but because of the stationarity of the energy functional with
respect to
$\nam
\sigma$, this affects $F_{\rm J}$ only in the order $\propto
D^4/\xi_0$, which is negligible for a pinhole.  Therefore we can use
the order parameter profiles calculated for a planar wall. 

Determining the suppression of the order parameter $\Deltavec$ at the 
wall is thus 
the first step needed for our calculation.
In the absence of mass currents and
magnetic scattering it is sufficient to consider the parametrization
\begin{equation} \label{e.gap} 
\Deltavec^{(0)}(\kvec,z)=
[\Delta_{\perp}(z)\zvec\zvec+
\Delta_{\parallel}(z)(\xvec\xvec+\yvec\yvec)]\cdot\kvec,
\end{equation} where $\zvec$ is perpendicular to the wall.  The gap
functions $\Delta_\perp(z)$ and $\Delta_\parallel(z)$, which are real
valued, are calculated self-consistently as explained in
Ref.\ \cite{zhang}. For $\nam{g}$ we use the ``randomly oriented
mirror'' (ROM) boundary condition at a specular or a diffusive surface
\cite{rom}. The numerical calculation of $\nam g$ is described in Sect.
\ref{s.pin}. Our results for $\Delta_\perp(z)$ and
$\Delta_\parallel(z)$ are similar to those found previously with ROM
and other surface models 
\cite{zhang,rom,verditzK,buchholtzrainer,buchholtz,krs,kopnin}.
In order to incorporate different phases and spin-orbit rotations on
the two sides we write
\begin{equation} \label{e.genedelta}
\Deltavec(\kvec,z)=\left\{\begin{array}{ll}
\exp(\iu \phi^L)\tensor{R}^L\!\cdot \Deltavec^{(0)}(\kvec,z) &
{\textrm{for}}~  z < 0 
\\
\exp(\iu \phi^R)\tensor{R}^R\!\cdot \Deltavec^{(0)}(\kvec,z) &
{\textrm{for}}~  z > 0 
\\
\end{array} \right. .
\end{equation}
The thin wall is located at $z=0$ and $\Deltavec^{(0)}$ (\ref{e.gap}) is
assumed to be symmetric:
$\Delta_\perp(-z)=\Delta_\perp(z)$ and 
$\Delta_\parallel(-z)=\Delta_\parallel(z)$.

\section{Energy functional} \label{s.ene}

Here we  derive a general quasiclassical expression for the Josephson
coupling energy in a pinhole.  
The derivation follows closely the lines of 
a quasiclassical treatment of impurities in $^3$He or superconductors
\cite{ions,pinning}. We start from an expression for the energy
difference between states with one impurity and no impurity, $\nam V$
being the impurity potential \cite{sr,free}. For a small spatial range
of $\nam V$, the self-energy $\nam \Sigma$ can be assumed to be unchanged
by it, and we get \cite{ions}
\begin{equation} \label{e.enefu}
\delta\Omega^{\rm tot}=-\frac{1}{2} 
\Tr [\ln(-\nam G_0^{-1}+\nam \Sigma+\nam
V)-\ln(-\nam G_0^{-1}+\nam \Sigma)]. 
\end{equation}
The trace operation $\Tr$ 
is defined as
\begin{equation} \label{e.trace}
\Tr \nam F  =
\kB T \sum_{m} \int\frac{\upd^3k}{(2\pi)^3} {\Tr}_4
\nam F(\vecb k,\vecb k,\epsilon_m), 
\end{equation}
where $\Tr_4$ denotes the trace of the $4\times 4$ Nambu
matrix $\nam F$. To eliminate the logarithm, we may apply some form of 
the ``$\lambda$-trick'' \cite{fetterwalecka}. We choose to integrate 
over the strength of
$\nam V$ by making the substitution $\nam V\rightarrow\lambda\nam V$
and writing
\begin{eqnarray} \label{e.lambdatrick}
\delta\Omega^{\rm tot}&=& 
\frac{1}{2}\Tr\int_0^1\frac{\upd\lambda}{\lambda}
(\nam G_0^{-1}-\nam\Sigma-\lambda\nam V)^{-1}\lambda \nam V \nonumber\\
&=&\frac{1}{2}\Tr\int_0^1\frac{\upd\lambda}{\lambda}\nam G_{1}\nam
T_{\lambda}.
\end{eqnarray}
The latter equality follows from a formal application of the
$t$-matrix equation $\nam T_{\lambda}=\lambda \nam V+\nam
T_{\lambda}\nam G_{1}\lambda\nam V$ and the relation $\nam G=\nam
G_{1}+\nam G_{1}\nam  T_{\lambda}\nam G_{1}$. Here $\nam G=(\nam
G_0^{-1}-\nam\Sigma-\lambda\nam V)^{-1}$  gives the full propagator in
the presence of an impurity scattering potential. The intermediate
Green's function $\nam G_{1}$ corresponds to a quasiclassical $\nam
g_{1}$ which has no discontinuity at the impurity. Equation
(\ref{e.lambdatrick}) is now in a form where the propagator can be
$\vert{\bf k}\vert$-integrated directly. However, to avoid a divergence
in the Matsubara summation, we have to subtract from
(\ref{e.lambdatrick}) the normal-state contribution
$\delta\Omega^{\rm N}$,  obtained by setting
$\nam\Sigma=\nam\Sigma^{\rm N}$ in
(\ref{e.enefu}). We define 
$\delta\Omega=\delta\Omega^{\rm tot}-\delta\Omega^{\rm N}$ and
transform this to the quasiclassical form
\begin{eqnarray} \label{e.quasifree}
\delta\Omega =\frac{1}{2}&&\efront \sum_{m} \int
\frac{\upd\Omega_\kvec}{4\pi} \int_0^1 \frac{\upd\lambda}{\lambda}
\nonumber\\
\times{\Tr}_4&&[\nam g_1(\kvec,\rvec_{\textrm{imp}},\epsilon_m)
\nam t_{\lambda}(\kvec,\kvec,\epsilon_m)\nonumber\\
&&-\nam g^{\rm N}_1(\kvec,\rvec_{\textrm{imp}},\epsilon_m)
\nam t^{\rm N}_{\lambda}(\kvec,\kvec,\epsilon_m)],
\end{eqnarray}
where $\rvec_{\textrm{imp}}$ is the location of the impurity and
$\nam t_\lambda(\kvec,\kvec,\me)$ is the forward-scattering part of the
$t$ matrix, which is obtained by solving
\begin{eqnarray} \breve t_\lambda(\kvec,\kvec',\epsilon_m)&=&
\lambda\breve v(\kvec,\kvec') +\lambda\pi N(0)\int
\frac{\upd\Omega_{\kvec''}}{4\pi}\breve v(\kvec,\kvec'')\nonumber\\
&&\times\breve g_1(\kvec'',\rvec_{\textrm{imp}},\epsilon_m)\breve
t_\lambda(\kvec'',\kvec',\epsilon_m). 
\label{e.tmatrixgen} \end{eqnarray}
The energy formula (\ref{e.quasifree}) is simpler to use than those in
Refs.\ \cite{ions,pinning}, since it does not involve an integration
over
$\rvec$. More importantly, $\nam g_1$ is constant in the $\lambda$
integration.

Now we specialize the above approach to the pinhole problem. The
coupling energy $F_{\rm J}$ we wish to know is, by definition, the
difference in energies between an open pinhole and a blocked pinhole.
It should be irrelevant how the hole is blocked as long as the
transmission of quasiparticles is prevented: changing the type of
blockage should only change some constant terms in the energy, which
do not depend on the soft degrees of freedom in (\ref{e.form2}). We
choose to block the pinhole by a surface just on the left hand side of
it (Fig.\ \ref{f.trajectory}). This surface is now considered as the
``impurity'' in Eqs.\ (\ref{e.enefu}-\ref{e.tmatrixgen}). The  coupling
energy is then equal to $\delta\Omega$ (\ref{e.quasifree}) except for a
minus sign:
$F_{\rm J}=-\delta\Omega$. The intermediate $\nam
g_1$ is the exact propagator that is
calculated for an open pinhole, and we drop the subindex 1 from here on.
 
The simplest choice for the blocking wall is a specularly
scattering surface. It corresponds to a delta-function scattering
potential ${\mathcal{V}}\delta(\hat{\bf l}\cdot{\rvec})$ 
for the (infinitesimal piece of) flat surface $\hat{\bf l}\cdot\rvec=0$
in the limit ${\mathcal{V}}\rightarrow\infty$.
The $t$ matrix of this type of
impurity is of particularly simple form
\cite{buchholtzrainer,sr} 
\begin{eqnarray} \label{e.tmatrix}
\nam
t_{\lambda}&&(\kvec,\kvec',\epsilon_m)=\nonumber\\
&&\frac{2\vF|\kvec\cdot\hat{\bf
l}|\lambda {\mathcal{V}}\upd A\delta^{(2)}_{{\bf k}_{\parallel},
{\bf k}'_{\parallel}}}
{2\vF|\kvec\cdot\hat{\bf l}|-\lambda{\mathcal{V}}[
\nam g(\kvec,\rvec_{\textrm{imp}},\epsilon_m)+
\nam g(\ul\kvec,\rvec_{\textrm{imp}},\epsilon_m)]},
\end{eqnarray}
where $\upd A$ is an area element ($\lambda_F^2 \ll\upd A \ll\xi_0^2$) of the blocking piece of wall 
with normal $\hat{\bf l}$. The component of $\kvec$ parallel to this
wall is denoted by
${\bf k}_{\parallel}=\kvec-(\kvec\cdot\hat{\bf l})\hat{\bf l}$, and 
$\ul\kvec=\kvec-2(\kvec\cdot\hat{\bf l})\hat{\bf l}$ is 
the mirror-reflected direction. 
We insert $\nam t_{\lambda}$ (\ref{e.tmatrix}) into
$\delta\Omega$ (\ref{e.quasifree}) and integrate over the blocking
surface. Performing also the
$\lambda$-integration, we find
\begin{eqnarray}
F_{\rm
J}(&&{\mathcal{V}})=-\frac{1}{2}\efront\int\upd A \sum_{m}
\int 
\frac{\upd\Omega_\kvec}{4\pi}|\kvec\cdot\hat{\bf l}|{\Tr}_4 \nonumber\\
&&\times\ln
\frac{2\vF|\kvec\cdot\hat{\bf l}|-{\mathcal{V}}[
\nam g(\kvec,\rvec_{\textrm{imp}},\epsilon_m)+
\nam g(\ul\kvec,\rvec_{\textrm{imp}},\epsilon_m)]}
{2\vF|\kvec\cdot\hat{\bf l}|+2\iu{\mathcal{V}}\nam \tau_3
\Sgn(\epsilon_m)}
\end{eqnarray}
where the normal-state propagator $\nam g^{\rm
N}=\iu\nam\tau_3\Sgn(\me)$ was used.
Next we take the limit $\mathcal{V}\rightarrow\infty$ and
use the general properties 
${\Tr}_4 \ln \nam A = \ln {\Det}_4\nam A$ and ${\Det}_4\nam A\nam 
B={\Det}_4\nam A {\Det}_4\nam B$, noting further that 
${\Det}_4[\iu\nam\tau_3\Sgn(\me)]=1$.
All trajectories that do not pass through the hole in the absence of
the blockade can be neglected, and thus the integration over the
surface $A$ can be transformed to one over the cross section (of area
$A_o$) of the hole.
This leads to the general pinhole energy functional 
\begin{eqnarray} \label{e.pinene}
F_{\textrm{J}}&&=
-\frac{1}{2}A_o\efront{\sum_{m}} 
\int\frac{\upd\Omega_\kvec}{4\pi}
|\hat k_z| \nonumber\\
&&\times\left\langle\ln\left\{{\Det}_4 \frac{1}{2}
[\nam g(\kvec,{\bf 0}^L,\epsilon_m)+
\nam g(\ul\kvec,{\bf 0}^L,\epsilon_m)]\right\}\right\rangle,
\end{eqnarray}
where $\hat k_z=\kvec\cdot\zvec$, 
the brackets $\langle\dots\rangle$ denote average over the trajectories (at
fixed $\kvec$) that hit the area $A_o$ of the
pinhole ($A_o=\pi D^2/4$ for a circular hole), and
${\bf 0}^L$ denotes the location immediately on the left hand side of
the hole.

The coupling energy (\ref{e.pinene}) depends on the shape of the
blocking wall only through the directions of the reflected momenta
$\ul\kvec=\kvec-2(\kvec\cdot\hat{\bf l})\hat{\bf l}$. Because the
surface can be chosen in different ways, there is a lot of
freedom in choosing the reflected
momenta. A particularly simple choice is 
\emph{retroreflection}: 
$\ul\kvec=-\kvec$. This can be achieved, for example, by a
semispherical blocking surface of radius $R$ satisfying $\xi_0\gg R\gg
D$, centered at the pinhole. This choice can simplify
practical calculations considerably, since one can use
symmetries effectively. After this one may, for example, calculate the
determinant by using 
${\mathrm{Det}}_4(\nam{g}_1+\nam{g}_2)
=[{\mathrm{Det}}_4(-2\nam{1}+\{\nam{g}_1,\nam{g}_2\})]^{1/2}$.

The choice to block the surface on the left hand side was arbitrary and
the blocking could equally well be done on the right hand side. 
The reason for not
blocking in the middle of the hole is that there may be scattering
taking place there. For example, a quasiparticle may hit the wall 
inside of
the hole and be deflected (Fig.\ \ref{f.trajectory}). In general this
causes 
$\breve g$ to be discontinuous at the hole, and in order to give an
unambiguous prescription for $F_{\rm J}$ (\ref{e.pinene}), one has to
specify one of the sides. If there is no scattering in the pinhole
($W/D=0$), the propagator is continuous and the energy functional 
can be evaluated in the middle of the hole. In
the absence of scattering also the average $\langle\dots\rangle$ in
$F_{\textrm{J}}$ (\ref{e.pinene}) is trivial and can be dropped.

Apart from the assumption of a pinhole aperture, the energy functional
(\ref{e.pinene}) is very general. Below we shall apply it for the
special case of $^3$He-B. 

\section{Pinhole calculation}\label{s.phcalc}

\subsection{Trajectories} \label{s.traj}

For the case $W/D=0$ and no scatterers localized within the pinhole, 
all trajectories hitting the orifice are directly transmitted. For the
case of finite $W/D$ we consider a model which is based on
the ROM boundary condition \cite{rom}. 

The ROM model assumes that the surface consists of microscopic pieces
of randomly oriented mirrors. Therefore, any trajectory hitting the
surface is simply deflected into another direction and the physical
propagator is continuous along it. When this process is averaged over
a length scale which is large compared to the size of the mirrors
($\lambda_{\rm F}\ll$ mirror size $\ll W$, $D$), one obtains a
probability distribution for the scattering from one
direction to another. Consider  a quasiparticle coming out of the
pinhole in direction $\kvec$. The probability density that it entered
the hole from direction $\kvec'$ can be written as
\begin{equation} \label{e.scatprob}
W_{\kvec,\kvec'}=p(\kvec)\delta_{\kvec,\kvec'}+
(1-p(\kvec))w_{\kvec,\kvec'}.
\end{equation} 
Here $p(\kvec)$ is the probability for direct
transmission, and $w_{\kvec,\kvec'}$ is the scattering distribution
obtained by averaging over the surface of the pinhole which is visible
from direction $\kvec$. 
For given $\kvec$ the distributions are normalized according to 
$\int(\upd\Omega_{\kvec'}/4\pi) W_{\kvec,\kvec'}=1$ and
$\int(\upd\Omega_{\kvec'}/4\pi) w_{\kvec,\kvec'}=1$, 
where the integrals are over \emph{all} directions (backscattering is 
also possible). For a circularly cylindrical aperture of diameter $D$
in a wall of thickness $W$ one finds a simple form for the transmission
function $p(\kvec)$:
\begin{equation} \label{e.probability}
p(\vartheta)=\left\{\begin{array}{ll}
\frac{2}{\pi}
(\gamma-\cos\gamma\sin\gamma) & {\textrm{for}} 
        ~  \vartheta<\arctan(D/W)\\
0 & {\textrm{for}} ~  \vartheta>\arctan(D/W),\\
\end{array} \right.   
\end{equation}
where $\gamma=\arccos(W/D)\tan\vartheta$ and $\vartheta$ is the polar
angle of the trajectory ($\kvec\cdot\zvec=\cos\vartheta$).

Calculating the $w_{\kvec,\kvec'}$ for a general type of ROM surface
and a finite ratio $W/D$ is difficult, since one has to consider 
multiple
scattering. We shall restrict to a cylindrical aperture and start with
fully diffusive, rough walls.  In this case one can in principle expand
$w_{\kvec,\kvec'}=P(\kvec|\kvec')=
\int\upd^2r_1 P_{\textrm{out}}(\kvec|r_1)P_{\textrm{in}}(r_1|\kvec')$
$+\int\upd^2r_1\int\upd^2r_2P_{\textrm{out}}(\kvec|r_2)
P(r_2|r_1)P_{\textrm{in}}(r_1|\kvec')$
$+\cdots$, where the $r_i$ parametrize positions on the surface of the
cylinder. For given outcoming direction $\kvec$ the function 
$P_{\textrm{out}}(\kvec|r)$ gives the distribution in $r$ for its
origin, whereas the function $P_{\textrm{in}}(r|\kvec')$ gives the 
distribution in incoming $\kvec'$, given a position of impact
$r$. The diffuse intermediate scatterings follow a ``Markovian''
process, so that the distribution $P(r_2|r_1)$ is independent of the
incoming  direction. The formal expansion parameter here is $W/D$ and
the first term in $w_{\kvec,\kvec'}$ is of zeroth order in it, corresponding to a single
scattering event. The second term has one intermediate scattering and
is thus of first order, and so on. In the limit $W/D\rightarrow 0$ only
the zeroth order term remains, and the functions $P_{\textrm{in}}$ and
$P_{\textrm{out}}$ are assumed to approach simple ``cosine laws'':
$P_{\textrm{in}}(r|\kvec')\propto|\svec(r)\cdot\kvec'|$ for 
$\svec(r)\cdot\kvec'<0$ and
$P_{\textrm{out}}(\kvec|r)\propto|\kvec\cdot\svec(r)|$ for
$\kvec\cdot\svec(r)>0$,  where $\svec(r)$ is the surface normal at
position $r$.  Upon normalization and insertion into the expansion for
$w_{\kvec,\kvec'}$ one finds
\begin{equation} \label{e.ddistr}
w_{\kvec,\kvec'}=\sin\vartheta_{\kvec'}
(\sin\chi_{\kvec,\kvec'}-\chi_{\kvec,\kvec'}\cos\chi_{\kvec,\kvec'}),
\end{equation} where 
$\chi_{\kvec,\kvec'}$
is the difference in incoming and outgoing azimuthal angles and 
$\vartheta_{\kvec'}$ is the incoming polar angle. This distribution is
largest for angles near $\chi_{\kvec,\kvec'}=\pi$ and
$\vartheta_{\kvec'}=\pi/2$, i.e., for scattering into directions close
to the plane of the surface.

As a second case, consider the possibility of specular scattering in
the pinhole.  Then a $\kvec'$ directional quasiparticle hitting the
surface at position $r$ will reflect into the direction 
$\ul{\kvec}'=\kvec'-2(\kvec'\cdot\svec(r))\svec(r)$.  In this case the
previous distribution functions 
$P_{\textrm{out}}, P_{\textrm{in}}$ and $P$ have to be generalized a
bit to take into account the non-Markovian character of the
scattering, but in the limit $W/D\rightarrow0$ no problems will arise.
Since $\svec$ is in the $xy$ plane, we necessarily have 
$\kvec\cdot\zvec=\kvec'\cdot\zvec$, and
a similar calculation as for the  diffusive case gives
\begin{equation} \label{e.sdistr}
w_{\kvec,\kvec'}=\pi(\sin\vartheta_{\kvec})^{-1}
\delta(\vartheta_{\kvec}-\vartheta_{\kvec'})
\sin(\chi_{\kvec,\kvec'}/2).
\end{equation}
More refined distributions could be obtained by taking into account
higher-order terms in the expansion, but doing so analytically would be
difficult. In Ref.\ \cite{rainerlee} these were briefly discussed in the
case of a long pore with specular walls.

\subsection{Propagator} \label{s.num}

Here we describe briefly the method used to generate the 
propagators. The idea is to calculate the propagators numerically only
for $\Deltavec^{(0)}$ (\ref{e.gap}), and to obtain analytically the
dependence of the true propagator on the soft degrees of freedom in
$\Deltavec$ (\ref{e.genedelta}). The matching of the left and
right solutions at the pinhole is most conveniently done with the
``multiplication trick'' \cite{zhang,rom,pinning}. There one first
calculates two unphysical solutions $\nam{g}_<$ and $\nam{g}_>$ of the
Eilenberger equation (\ref{e.eil}) and $\nam{g}$ is
constructed using
\begin{equation} \label{e.scatmult}
\nam{g}(\kvec,\rvec,\me)=
\iu \mathop{\rm Sgn}(\hat k_z)
\frac{[\nam{g}_<(\kvec,\rvec,\me),\nam{g}_>(\kvec,\rvec,\me)]}
{\{\nam{g}_<(\kvec,\rvec,\me),\nam{g}_>(\kvec,\rvec,\me)\}}.
\end{equation}
Here we denote by $\nam g_<$ and $\nam g_>$ the solutions
decaying exponentially towards left ($z=-\infty$) and right
($z=+\infty$), respectively, independently of the direction of $\kvec$.

We rewrite the propagator components as 
$g=c+d$, $\gvec=\cvec+\dvec$, $f=a+b$, $\fvec=\avec+\bvec$ and
$\td g=c-d$, $\td \gvec=\cvec-\dvec$, $\td f=a-b$, 
$\td \fvec=\avec-\bvec$.  In terms of these, the
Eilenberger equation decouples conveniently into three independent
blocks of linear, first-order differential equations which are
numerically more convenient to handle \cite{zhang}. The first task is
to find the unphysical solutions consisting of components $a$, $b$,
and ${\bf c}$.  
For the real valued order parameter
$\Deltavec^{(0)}$, the unphysical propagator components can be chosen
such that $a$ and $b$ are real and
$\cvec$ is purely imaginary. The unphysical block of equations
\cite{zhang}  then becomes 
\begin{eqnarray} \label{e.upblock}
\me b+\frac{1}{2}\hbar\vF\partial_ua&=&0 \nonumber\\
\me a+\Deltavec^{(0)}\cdot\cvecim+
	\frac{1}{2}\hbar\vF\partial_ub&=&0 \nonumber\\
-\nuvec\times\cvecim+\Deltavec^{(0)} b+
	\frac{1}{2}\hbar\vF\partial_u\cvecim&=&\bf 0. 
\end{eqnarray}
Here $u$ is the coordinate along an arbitrary trajectory
$\rvec=\rvec_0+u\kvec$, and we fix $u=0$ at the wall ($z=0$). 
In accordance with (\ref{e.scatmult}), the exponential
solutions of (\ref{e.upblock}) which go through the pinhole are 
denoted by  
$\nam g_<^{(0)}$ and $\nam g_>^{(0)}$, respectively. These are the
solutions that are naturally obtained by integrating from the
bulk towards the wall on $L$ and $R$ sides. Because of symmetries we
only need to calculate $\nam g_<^{(0)}$, and we do this 
using fourth-order Runge-Kutta method. We introduce a
short-hand notation for the numerically calculated quantities
\begin{eqnarray} \label{e.shorthand}
A(\kvec,\me)&\equiv& a^{(0)}_<(\kvec,u=0,\me)\nonumber\\
B(\kvec,\me)&\equiv& b^{(0)}_<(\kvec,u=0,\me)\nonumber\\
\Cvec(\kvec,\me)&\equiv& \cvec^{(0)}_<(\kvec,u=0,\me).
\end{eqnarray}
These were evaluated for several directions of
$\kvec$, whose polar angles $\vartheta$ were chosen so that angular
integrations in (\ref{e.progcurrent}) and (\ref{e.final}) could be 
carried
out using the Gaussian quadrature, usually with 32 points in the range
$\cos\vartheta=-1\ldots1$.  (Due to the symmetry of the integrands, 
only values for 
$\hat k_z>0$ actually need to be considered.)
The number of (positive) Matsubara energies 
$\me=\pi\kB T(2m+1)$ was between 10 to around 100 depending on the
temperature. 
Above we assumed $\Deltavec^{(0)}$ and $\nuvec$ to be already known.
The method for their self-consistent calculation is the same as above,
except that also the solutions diverging away from the wall have to be
calculated. The initial conditions for these were obtained using the 
specular reflection or the ROM boundary condition.

The functions $\nam g_>^{(0)}$ can be obtained by using the relations
\begin{eqnarray} \label{e.extrasymm}
a_>^{(0)}(\kvec,u,\me) & = &+a_<^{(0)}(\kvec,-u,\me) \nonumber\\
b_>^{(0)}(\kvec,u,\me) & = &-b_<^{(0)}(\kvec,-u,\me) \nonumber\\
\cvec_>^{(0)}(\kvec,u,\me) & = &+\cvec_<^{(0)}(\kvec,-u,\me),
\end{eqnarray}
which are based on the symmetry 
\begin{equation} \label{e.psymm}
\Deltavec^{(0)}(\kvec,u)=\Deltavec^{(0)}(\kvec,-u).
\end{equation} 
From the solutions $\nam g_<^{(0)}$ for $\Deltavec^{(0)}$ we
get the solutions $\nam g_<$ for the general $\Deltavec$
(\ref{e.genedelta}) on the $L$ side by forming the linear
combinations  
\begin{eqnarray} \label{e.generals}
a_< & = &
a_<^{(0)}\cos\phi^L+\iu b_<^{(0)}\sin\phi^L
\nonumber\\ b_< & = &
\iu a_<^{(0)}\sin\phi^L+b_<^{(0)}\cos\phi^L
\nonumber\\
\cvec_< & = & 
\hspace{1mm}\dyadic{R}^L\!\!\cdot\cvec_<^{(0)}.
\end{eqnarray}
The same equations hold on the $R$ side when $L$ is replaced by $R$ and
$<$ by $>$. 

The physical propagator at the pinhole ($\rvec=\rvec_0=\vecb 0$)
can now be obtained using Eqs.\ 
(\ref{e.scatmult}), (\ref{e.shorthand}), (\ref{e.extrasymm}), and
(\ref{e.generals}). For the case of deflected trajectories we have to
specify separately the momentum  
$\hat{\bf k}'$ on $L$ side and $\hat{\bf k}$ on $R$ side.
Only the transmitted trajectories ($\hat k_z\hat k_z'>0$) need to be
considered, and we get
\begin{eqnarray}
\avec_{\kvec,\kvec'}(\epsilon_m) & &~=
\iu~\nc_{\kvec,\kvec'}^{-1}[\Cvec_L'~ 
(\iu A\sin\frac{1}{2}\phi-B\cos\frac{1}{2}\phi) \nonumber \\
& &+(\iu A'\sin\frac{1}{2}\phi-B'\cos\frac{1}{2}\phi)\Cvec_R] 
\nonumber\\
\bvec_{\kvec,\kvec'}(\epsilon_m) & &~= 
\iu~\nc_{\kvec,\kvec'}^{-1}[\Cvec_L'~
(A\cos\frac{1}{2}\phi-\iu B\sin\frac{1}{2}\phi) \nonumber \\
& &-(A'\cos\frac{1}{2}\phi-\iu B'\sin\frac{1}{2}\phi)\Cvec_R] 
\nonumber\\ 
d_{\kvec,\kvec'}(\epsilon_m)& &~= \nonumber \\
\iu~\nc_{\kvec,\kvec'}^{-1}& &[\iu
(AA'+BB')\sin\phi-(AB'+A'B)\cos\phi] \nonumber\\[1.1mm]
\dvec_{\kvec,\kvec'}(\epsilon_m) & &~=-\nc_{\kvec,\kvec'}^{-1} 
\Cvec_L'\times\Cvec_R,
\label{e.physprop4}
\end{eqnarray} 
where $\Cvec_{L,R}=\tensor{R}^{L,R}\!\cdot\Cvec$, and primes denote
values corresponding to direction $\kvec'$. The normalization
constant is given by
\begin{eqnarray} \label{e.nckk}
\nc& &_{\kvec,\kvec'}(\epsilon_m)
=-\nc^*_{-\kvec,-\kvec'}(\epsilon_m)=\mathop{\rm Sgn}(\hat k_z)\\ &
&\times[-(AA'+BB')\cos\phi+\iu(A'B+AB')\sin\phi+\Cvec_L'\!\cdot\Cvec_R].
\nonumber
\end{eqnarray}
For the case of direct transmission ($\kvec'=\kvec$) these
expressions  simplify considerably.

\subsection{Currents and coupling energy} \label{s.josenergy}

As an application of the above results, consider the Josephson current
in the pinhole.  Using the general symmetries for propagators, 
the mass current density (\ref{e.genecurrent}) can be written in terms 
of $\re d$ alone. The total current is then given by $\re
d_{\kvec,\kvec'}$  (\ref{e.physprop4}) integrated over the distribution
(\ref{e.scatprob}) of trajectories:
\begin{eqnarray}  \label{e.progcurrent}
J=&&A_o\jfront \nonumber\\ 
&&\times\sum_{m}\int \frac{\upd\Omega_\kvec}{4\pi}
\hat k_z
\int\frac{\upd\Omega_{\kvec'}}{4\pi}W_{\kvec,\kvec'}
\re d_{\kvec,\kvec'}(\epsilon_m).
\end{eqnarray}

In the case of direct transmission only
($W/D=0$ or $W_{\kvec,\kvec'}=\delta_{\kvec,\kvec'}$), 
one can apply trigonometric identities to put 
$d(\kvec,{\bf 0},\epsilon_m)=d_{\kvec,\kvec}(\epsilon_m)$ in the form
\begin{eqnarray} \label{e.sophisticated}
d(&&\kvec,\vecb 0,\epsilon_m)=\mathop{\rm Sgn}(\hat k_z)\nonumber\\ 
&&\times\frac{1}{4} 
\sum_{\sigma=\pm1} 
\frac{(B^2-A^2)\sin(\phi+\sigma\zeta)+2\iu A B}
{A^2\sin^2[\frac{1}{2}(\phi+\sigma\zeta)]+
B^2\cos^2[\frac{1}{2}(\phi+\sigma\zeta)]
},
\end{eqnarray}
where $\zeta(\kvec,\epsilon_m)$ is defined by 
$\Cvec_L\cdot\Cvec_R=\Cvec^2\cos\zeta$.
The real part of (\ref{e.sophisticated}) is now equivalent to 
equation (1) in Ref.\ \cite{yippi}, but more general. The quantities 
$\phi\pm\zeta(\kvec,\epsilon_m)$ are the effective phase differences 
experienced
by quasiparticles with different spin projections along the axis 
$\Cvec_L\times \Cvec_R$
$\propto\Deltavec_L(\kvec)\times \Deltavec_R(\kvec)$. 
In the special case that
$\Deltavec$ is assumed constant, i.e., unsuppressed at the walls, the
Matsubara summation can be done analytically  and one obtains the same
result for $J$ as in Ref.\ \cite{yippi}.

For the Josephson coupling energy (\ref{e.pinene}) we find
\begin{eqnarray} \label{e.final}
F_{\textrm{J}}=&&
\frac{1}{2}A_o\efront\sum_{m} \int\frac{\upd\Omega_\kvec}{4\pi}
|\hat k_z|\\
&&\times 
\int\frac{\upd\Omega_{\kvec'}}{4\pi}W_{\kvec,\kvec'}
\left\{\ln|\nc_{\kvec,\kvec'}(\epsilon_m)|^{2}-
\ln[4(AA')^2]\right\}. \nonumber
\end{eqnarray}
In the first term there appears the squared modulus of the 
trajectory-invariant normalization constant (\ref{e.nckk}).
The second term has to be retained to have convergence in the
Matsubara summation. 

Consider again the direct-transmission case $W/D=0$.
In the Ginzburg-Landau limit $T\rightarrow T_{\rm c}$ we can 
verify the phenomenological form
(\ref{e.fj}) and calculate the parameters $\alpha$ and $\beta$. In this
limit the amplitude of 
$\Deltavec$ is small and since 
$|\Deltavec|^2\sim|\Cvec|^2=|A^2-B^2|$, we should have 
$|A^2-B^2|\ll A^2+B^2$, $A^2\approx B^2$ and $|s_{\kvec,\kvec}
|^2\approx4A^4$. 
It follows that the logarithm in the first term of
(\ref{e.final}) can be expanded to linear order to give (\ref{e.fj}),
where 
\begin{eqnarray} 
\alpha && =A_o\efront\!\int \frac{\upd\Omega_\kvec}{4\pi}
|\hat k_z|
\sum_m\frac{(\mathop{\textrm{Im}} C_z)^2}{A^2+B^2},
\nonumber\\
\beta && =\frac{1}{2}A_o\efront\!\int
\frac{\upd\Omega_\kvec}{4\pi} 
|\hat k_z|
\sum_m\frac{(\mathop{\textrm{Im}} C_{\rho})^2}{A^2+B^2}. \label{e.beta}
\end{eqnarray} 

\begin{figure}[!tb]
\begin{center}
\includegraphics[width=0.9\linewidth]{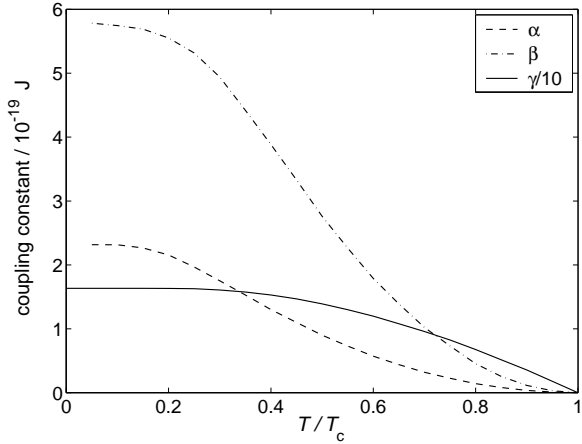}
\caption{Temperature dependence of the tunneling model parameters
$\alpha$, $\beta$ (\ref{e.beta}), and $\gamma$ (\ref{e.gammapar}) as
calculated for a diffusive wall and total open area $A_o=A\kappa$ where,
$A=3.8\cdot10^{-8}$ m$^2$, 
$\kappa=14.7\cdot10^{-4}$. The pressure is 0 bar, $W/D=0$, and
$F_i^a=0$ for $i=1$, 3, etc.}
\label{f.tconst}
\end{center}
\end{figure}

Figure \ref{f.tconst} shows the temperature dependence of the
tunneling parameters for a diffusive wall with $A_o$ chosen to match the
total open area of a coherent array of holes with dimensions as in
Ref.\ \cite{bistability}. Also shown is the textural rigidity parameter
$\gamma$ whose role is to be discussed below in Sect.\ \ref{s.synerg}.
Close to $T_{\rm c}$ the strength of the coupling, i.e., the parameters
$\alpha$ and $\beta$ go as
$\alpha,\beta\propto(1-T/T_{\rm c})^2$, whereas the rigidity 
$\gamma\propto(1-T/T_{\rm c})$. 

The Josephson current (\ref{e.progcurrent}) and energy (\ref{e.final})
were obtained completely independently.  It is essential to check that
they are consistent with each other. One can easily see that the
component $\re d_{\kvec,\kvec'}$ satisfies 
\begin{equation}
\re d_{\kvec,\kvec'}(\epsilon_m)
=\frac{1}{2}\mathop{\rm Sgn}(\hat k_z)
\frac{\partial}{\partial\phi}\ln|\nc_{\kvec,\kvec'} (\epsilon_m)|^2,
\end{equation} and hence the macroscopic current formula
(\ref{e.enederiv}) is exactly satisfied. As a further check of
the energy (\ref{e.final}) we can see that also the spin current formula
(\ref{e.plausible}) is satisfied. Using  $\Cvec'_L\cdot\Cvec_R=R^L_{\mu
i}R^R_{\mu j}C'_iC_j$ the  energy (\ref{e.final}) is seen to be a
function of the products
$R^L_{\mu i}R^R_{\mu j}$ and calculating the spin current from 
(\ref{e.plausible}) is thus possible. Writing also
$[\Cvec'_L\times\Cvec_R]_\gamma=\epsilon_{\alpha\beta\gamma}R^L_{\alpha
k}R^R_{\beta l}C'_kC_l$ in the propagator component (\ref{e.physprop4})
and using the quasiclassical spin current expression
(\ref{e.spincurrent}) it can be checked that the result for
$J_\gamma^{\rm spin}$  agrees  with the one obtained from
(\ref{e.plausible}). 

\section{Textural interactions} \label{s.surf}

As discussed in Sect.\ \ref{s.symm}, the Josephson effect in $^3$He
depends on the rotation matrices $R_{\mu i}^{L,R}$ on the two sides of
the weak link.  These matrices are determined by the
competition of a number of relatively weak bulk and surface
interactions, which lift the degeneracy of the
B phase order parameter (\ref{e.opa}). The equilibrium configuration is
found by minimizing a hydrostatic energy functional. We shall present
the hydrostatic theory to the extent needed here. For a recent review
see Ref.\ \cite{hydro}.

\subsection{Interactions and coupling constants}

The most important hydrostatic energy term arises from the dipole-dipole
interaction between the nuclear magnetic moments,
\begin{equation} \label{e.dipint}
F_{\textrm{D}}=8g_{\textrm{D}}\Delta^2\int\upd^3r(\frac{1}
{4}+\cos\theta)^2,
\end{equation} 
where $\theta$ is the rotation angle of the spin-orbit rotation $R_{\mu
i}(\nvec,\theta)$. The effect of $F_{\textrm{D}}$ is to fix $\theta$ to
the value
$\theta_0\approx104^\circ$ in the bulk liquid. 

There is no conflict between $F_{\textrm{D}}$ (\ref{e.dipint}) and
$F_{\textrm{J}}$ (\ref{e.form2}). Even if
$R_{\mu i}^{L,R}$ have their rotation angles fixed to
$\theta_0$, $F_{\textrm{J}}$ depends on their product $\psi_{ij}$ ---
also a rotation matrix --- and it can attain \emph{all} possible values
if $\nvec^L$ and
$\nvec^R$ are directed properly. Thus $\theta$ is not changed from 
$\theta_0$ by the Josephson coupling. The same applies to all surface
energies below. We therefore assume 
$F_{\textrm{D}}$ to be in its minimum everywhere and study only the
position-dependence of
$\nvec(\rvec)$, which is known as the texture.

In the absence of a magnetic field the dominant interaction
determining the texture near a wall is the surface-dipole interaction
\begin{equation} \label{e.fsd}
F_{\textrm{SD}}=\int_S\upd^2r[b_4(\svec\cdot\nvec)^4
-b_2(\svec\cdot\nvec)^2].
\end{equation} where $b_2$ and $b_4$ are positive coupling constants
and $\svec$ is the surface normal. There are usually many
walls with different orientations present and 
therefore there is a conflict between their orienting effects. This
leads to gradient (bending) energy
\begin{equation} \label{e.bge} 
F_{\textrm{G}}=
\int\upd^3r\left [
\lambda_{\textrm{G1}}
\frac{\partial R_{\alpha i}}{\partial r_i}
\frac{\partial R_{\alpha j}}{\partial r_j}+
\lambda_{\textrm{G2}}
\frac{\partial R_{\alpha j}}{\partial r_i}
\frac{\partial R_{\alpha j}}{\partial r_i} \right],
\end{equation}
which is related to spin currents in the bulk.
The gradient energy also has a surface part 
\begin{equation} \label{e.sge}
F_{\textrm{SG}}=\lambda_{\textrm{SG}}
\int_S\upd^2r \hat{s}_j R_{\alpha j}
\frac{\partial R_{\alpha i}}{\partial r_i}.
\end{equation}
In the presence of a magnetic field another surface interaction
becomes important:
\begin{equation} \label{e.sh}
F_{\textrm{SH}}=-d\int_S \upd^2r
(\vecb H\cdot \tensor{R}\cdot \svec)^2.
\end{equation}
There is also a bulk magnetic interaction
\begin{equation} \label{e.dh}
F_{\textrm{DH}}=-a\int \upd^3r(\nvec\cdot \vecb H)^2.
\end{equation}
The effect of stationary flow on the texture could be incorporated by 
the dipole-velocity interaction $F_{\textrm{DV}}$, but the flow
velocities in the experiment
\cite{bistability} are so small that $F_{\textrm{DV}}$ is negligible.

The values of the many coefficients appearing above are discussed in
Ref.\ \cite{hydro} in some detail. However, most of them have 
only been evaluated in the GL region so far. 
As a byproduct of our calculation 
of the surface order parameter, we can now extend the calculations of 
$b_2$, $b_4$ and $\lambda_{\textrm{SG}}$ to all temperatures.

\begin{figure}[!tb]
\begin{center}
\includegraphics[width=0.9\linewidth]{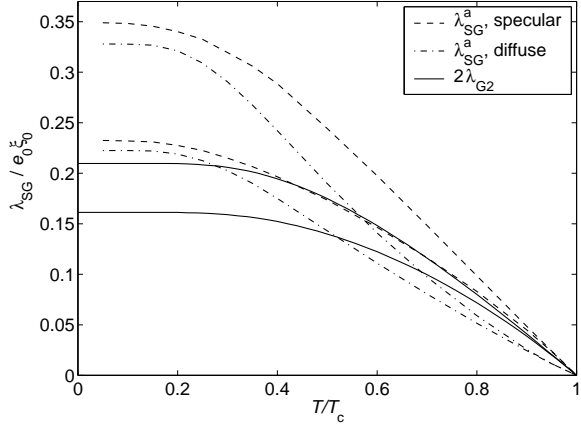}
\caption{Gradient energy parameters $\lambda_{\textrm{SG}}^a$ (dashed
line: specular wall, dash-dotted line: diffusive wall) and 
$2\lambda_{\textrm{G2}}$ (solid line) in the
weak coupling approximation. The upper curves are for $F_1^a=0$ and
the lower ones for $F_1^a=-1$. $F_3^a$ and higher order parameters are
assumed zero.
The unit is $e_0\xi_0$, where $e_0=\hbar\vF\NF\kB T_{\rm c}$.}
\label{f.lasgp}
\end{center}
\end{figure}

The surface gradient parameter consists of two contributions
$\lambda_{\textrm{SG}}=\lambda_{\textrm{SG}}^a+2\lambda_{\textrm{G2}}$.
The contribution $\lambda_{\textrm{SG}}^a$
is equal to 
$J^{\textrm{spin}}_{y,x}/L_y$, the current of $y$-directional spin
projection in the 
$x$ direction per unit length
in the $y$ direction  ($L_y$), calculated for the order
parameter $\Deltavec^{(0)}$. In terms of spin
current density ${\bf j}^{\textrm{spin}}_\gamma$
(\ref{e.spincurrent}) it is given by
\begin{equation}
\lambda_{\textrm{SG}}^a=\int_0^\infty\upd z
j^{\textrm{spin}{(0)}}_{y,x}(z).
\end{equation}
Figure \ref{f.lasgp} gives the temperature dependence of both terms
of
$\lambda_{\textrm{SG}}$ for the cases of a specular and a diffusive wall.
Close to $T_{\rm c}$ all of the parameters vanish linearly in $T-T_{\rm c}$. The
slopes of 
$\lambda_{\textrm{SG}}=\lambda_{\textrm{SG}}^a+2\lambda_{\textrm{G2}}$
agree with the GL results of Ref.\ \cite{hydro} for vapor pressure; at
low temperatures there is considerable deviation from the linear GL
behavior. We have calculated the results for $F_1^a=0$ and $F_1^a=-1$, 
the true value at vapor pressure being probably somewhere in between.
The change with $F_1^a$ is rather strong here, since
$\lambda_{\textrm{SG}}^a$ is directly related to the spin current.

\begin{figure}[!tb]
\begin{center}
\includegraphics[width=0.9\linewidth]{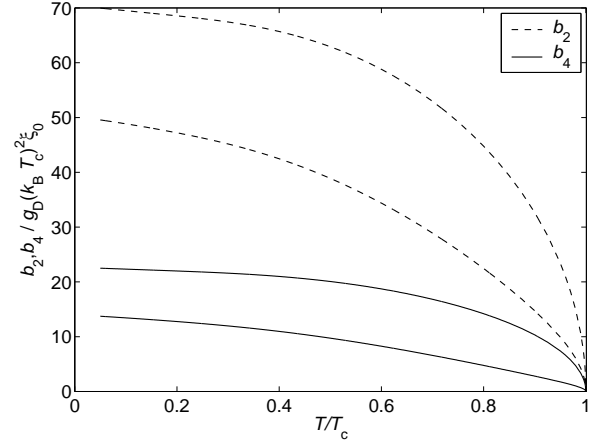}
\caption{Surface dipole interaction parameters $b_2$ (dashed lines)
and $b_4$ (solid lines) in the weak coupling approximation and with 
$F_1^a=0$. 
The lower of each pair is for a diffusive wall and the upper
for a specular wall. The unit of $b_2$ and $b_4$ is $g_{\textrm{D}}(\kB
T_{\rm c})^2\xi_0$.}
\label{f.bpar}
\end{center}
\end{figure}

The surface-dipole coupling constants $b_2$ and $b_4$ are given by
\cite{hydro}
\begin{eqnarray}
b_2&=&\frac{5}{4}g_{\textrm{D}}\int_0^\infty\upd z
(\Delta_\perp^2-6\Delta_\perp\Delta_\parallel+5\Delta_\parallel^2)
\nonumber\\ b_4&=&\frac{25}{8}g_{\textrm{D}}\int_0^\infty\upd
z(\Delta_\parallel-\Delta_\perp)^2
\end{eqnarray}
at all temperatures.
Figure \ref{f.bpar} shows these two plotted as a function of $T/T_{\rm c}$.
Both of them go to zero proportional to $(1-T/T_{\rm c})^{1/2}$ near $T_{\rm c}$.
In this region more accurate values can be obtained from the  GL
results \cite{hydro}, with which these coincide. Because $b_2>2b_4$,
$F_{\textrm{SD}}$ (\ref{e.fsd}) favors $\nvec$
perpendicular to the wall. The dependence 
of $b_2$ and $b_4$ on $F_1^a$ is much weaker than that of 
$\lambda_{\textrm{SG}}$, since the effect of the spin current comes
to play only through the order parameter. For $F_1^a=-1$ the values 
tend to be diminished 
by at most 5 percent from those shown in Fig.\ \ref{f.bpar}.

\subsection{Competing interactions and length scales}

In dealing with the pinhole model, we are mostly interested in the behavior of the texture 
at different magnetic fields near a flat surface. 
In this case there are three relevant
contributions to energy (per surface area $L^2$): 
(i) the surface-dipole energy, $F_{\textrm{SD}}/L^2\approx b_2$,
(ii) the surface-field energy, $F_{\textrm{SH}}/L^2\approx dH^2$, and
(iii) one related to the bending of the
texture when there is a uniform perturbation at the wall (caused by
$F_{\rm SH}$, for example) $(F_{\textrm{G}}+F_{\textrm{DH}})/L^2$
$\approx\sqrt{65\lambda_{\textrm{G2}}aH^2/8}\propto H$.
All of these have different field dependences, but their values turn 
out to coincide at $H\approx1$ mT. It is seen that at
fields $H\lesssim$ 1 mT the constant $F_{\rm SD}$ always dominates, 
and thus $\nvec$  will be aligned perpendicular to the surface.
For $H\gtrsim$ 1 mT the dominant surface interaction is $F_{\rm SH}$,
and the local texture is determided by its minima.

The conclusion about the relative magnitudes of $F_{\rm SD}$ and 
$F_{\rm SH}$ is true also in general. However, to determine the
texture at the surfaces in more complicated restricted geometries, the
gradient and other bulk energies have to be minimized together with
the surface energies. The surfaces can then be seen as perturbing the bulk
texture which would otherwise be uniform.
Associated with each hydrostatic interaction competing with the gradient
energy, there is a characteristic length scale that describes the scale
on which such local perturbations from uniformity will decay, or heal
\cite{smith,vollhardt}. The stronger the competing
interaction, the shorter is the corresponding healing length. Comparing these
healing lengths with the spatial scale $l\gg\xi_0$ of the container gives
qualitative information on the form of the texture.

Most importantly, in the presence of a magnetic field 
$F_{\rm DH}$ and $F_{\rm G}$ define a length
$\xi_{\rm H}=\sqrt{65\lambda_{\textrm{G2}}/(8aH^2)}\propto H^{-1}$ \cite{hydro}.
Comparison of $F_{\rm D}$ and $F_{\rm G}$ gives another one, the dipole
length $\xi_{\rm D}=\sqrt{\lambda_{\textrm{G2}}/g_{\textrm{D}}\Delta^2}\approx
10~\mu$m $\gg \xi_0$, related to possible
perturbations of the rotation angle $\theta$ from its equilibrium
value $\theta_0$. As stated above, we assume there to be no interactions
present which could force such perturbations, and therefore 
$\xi_{\rm D}$ plays no role. 
In any case, for most practical purposes
$\xi_{\rm H},l\gg\xi_{\rm D}$, and such a perturbation would decay
fast on the scale of $l$. 

For very small magnetic fields $\xi_{\rm H}\gg l$
and $\xi_{\rm H}$ is thus also not a relevant length scale.
The only important hydrostatic interactions in this case are 
the surface-dipole energy $F_{\rm SD}$ and the gradient energies
$F_{\rm G}$ and $F_{\rm SG}$.  Now the interesting question
is simply whether the walls of the container are separated by long enough
distances for $F_{\textrm{SD}}$ to be essentially minimized, or  small
enough distances for no textural variation to occur at all --- the
minimum configuration of 
$F_{\textrm{Gtot}}=F_{\textrm{G}}+F_{\textrm{SG}}$.  
An elementary estimation of the length scale $l$ at which
there is a transition from one behavior to another gives 
$l\propto\lambda_{\textrm{G2}}/(b_2-b_4)$, where the constant of
proportionality is of order unity. Since $b_4<b_2$, we can 
drop it and define a \emph{surface-dipole length} 
$\xi_{\textrm{SD}}\equiv\lambda_{\textrm{G2}}/b_{2}$.
Using the numerical values calculated above we get at zero pressure
\begin{equation} \label{e.xisd}
\xi_{\textrm{SD}}=\frac{\lambda_{\textrm{G2}}}{b_{2}}\approx
\left\{
\begin{array}{ll}
2.5~\textrm{mm}\quad &T\approx 0 \\
6.8(1-T/T_{\rm c})^{1/2}~\textrm{mm}\quad &T\approx T_{\rm c}.
\end{array} \right.
\end{equation}

\section{Isotextural Josephson effect} \label{s.pinres}

The Josephson energy (\ref{e.form3}) of a pinhole depends nontrivially
on three parameters: the phase $\phi$ and two parameters
describing $\psi_{ij}=R_{\mu i}^LR_{\mu j}^R$. In addition there is
dependence on the surface scattering (diffusive vs.\ specular), on the
temperature
$T/T_{\rm c}$ and on
$W/D$. Below we can present only some representative parts of this
parameter space. In this section we plot isotextural current-phase
relationships, where $\psi_{ij}$ is assumed constant in
$J(\phi,\psi_{ij})$. The possible $\phi$ dependence of $\psi_{ij}$ is
considered in Sec.\ \ref{s.synerg}.

The pinhole coupling has the maximal symmetry (\ref{e.form3})
independently of the shape of the hole as long as $W/D=0$. In terms of
$\nvec$ this means, for example, $F_{\rm J}(\phi,\nvec^L,\nvec^R)
=F_{\rm J}(\phi,\nvec^R,\nvec^L) =F_{\rm J}(\phi,-\nvec^L,-\nvec^R)$,
and similarly for the mass current. The mass current is given in units
of 
$J_0=2m_3\vF\NF\kB T_{\rm c}A_o$, where $A_o$ is the total open area of one
or more pinholes. In most cases only the phase differences in the range
$[0,\pi]$ are shown due to the symmetry
$J(2\pi-\phi)=-J(\phi)$. All plots are made for
$F_1^a=0$. Tests with
$F_1^a=-1$ show no qualitative differences and at most a few percent
quantitative difference in
$J(\phi)$ even at the lowest temperatures.

\subsection{Spin-rotation axes perpendicular to wall}

\begin{figure}[!tb]
  \begin{minipage}[t]{.99\linewidth}
    \centering \includegraphics[width=0.8\linewidth]{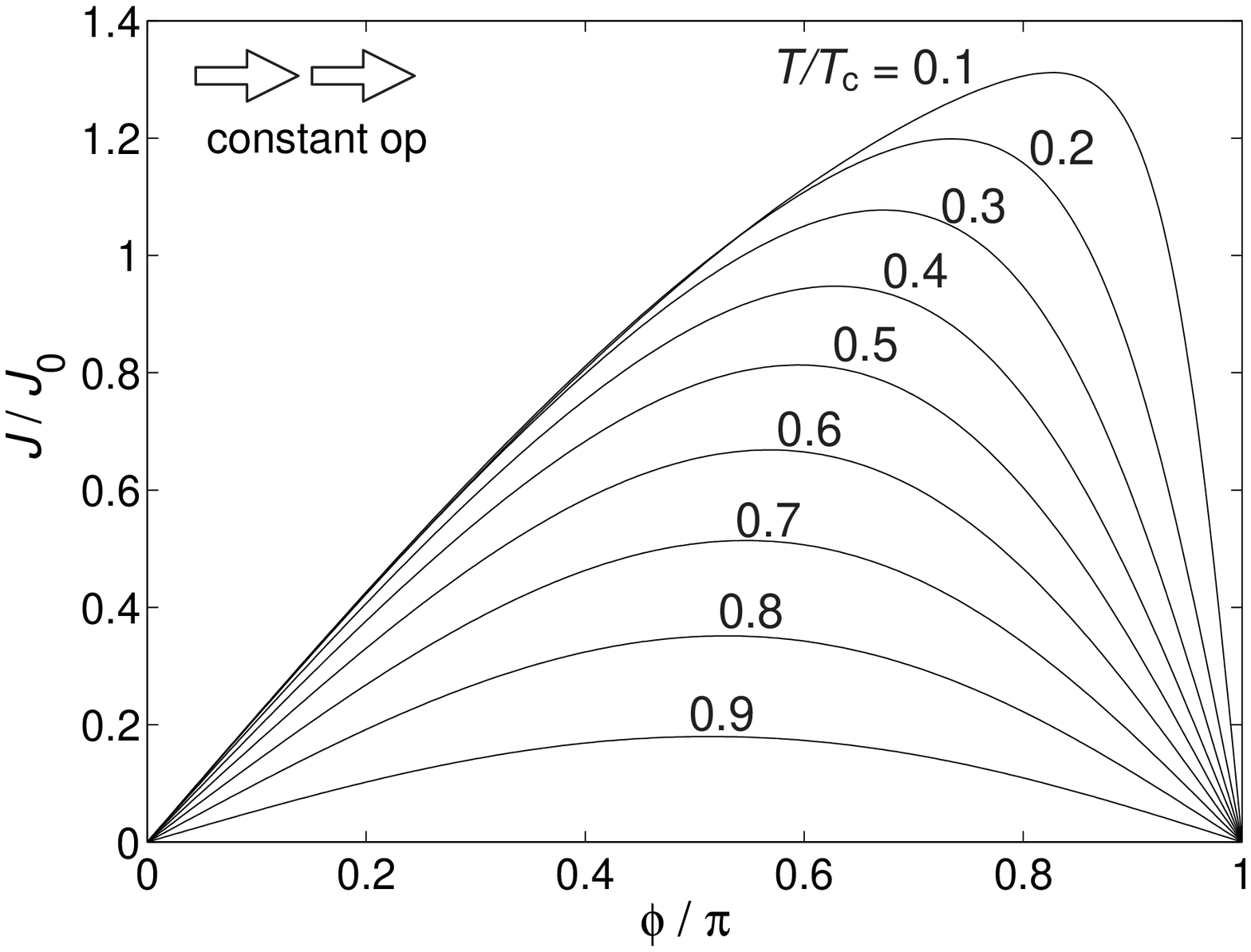}
  \end{minipage}
  \hfill
  \begin{minipage}[t]{.99\linewidth}
    \centering \includegraphics[width=0.8\linewidth]{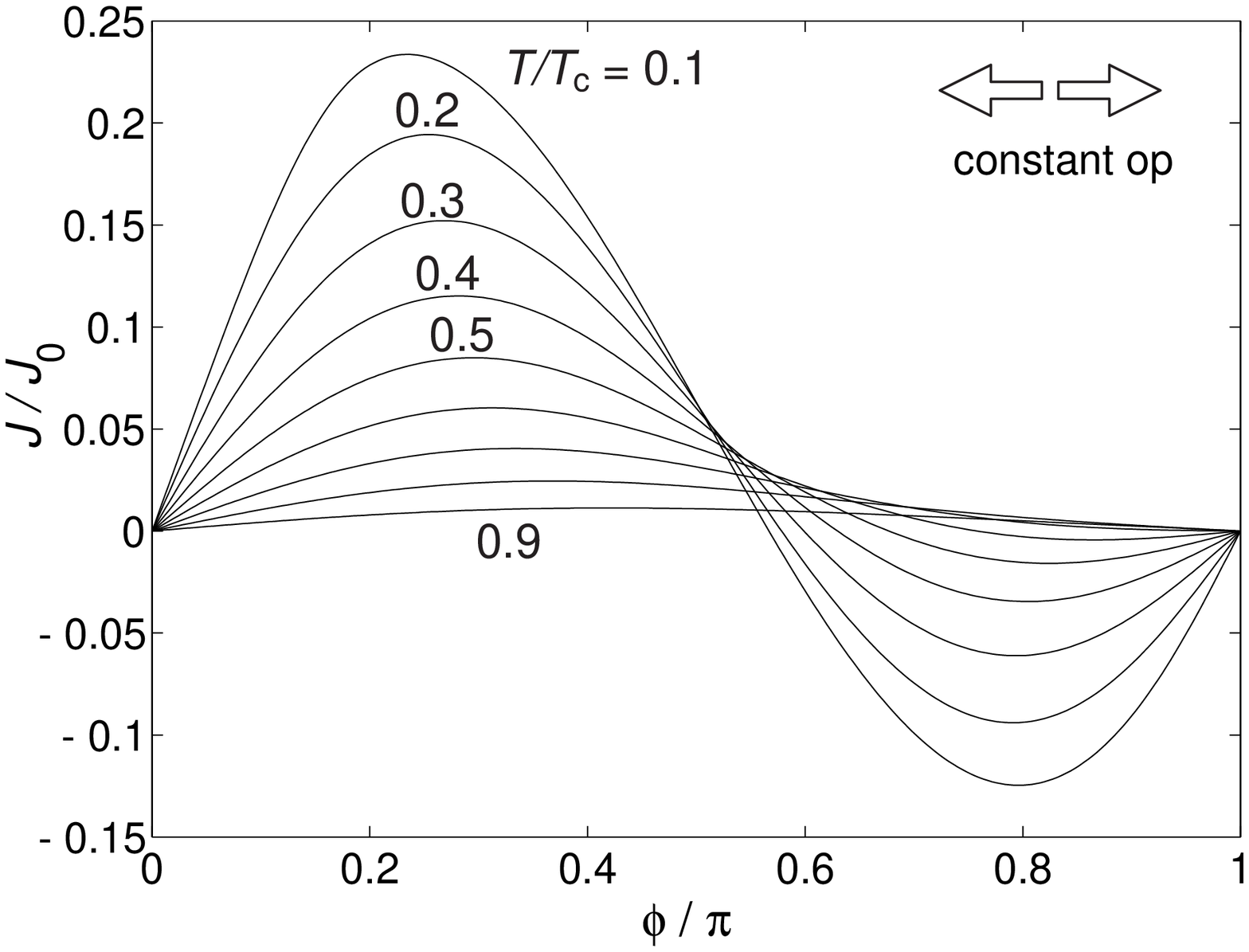}
  \end{minipage}
\caption{Isotextural current-phase relations for a pinhole in a wall with a constant
order parameter on both sides. The top panel corresponds to
$\nvec^L=\nvec^R$ and lower to $\nvec^L=-\nvec^R=\pm\zvec$. $W/D=0$.}
\label{f.kiphi}
\end{figure}

\begin{figure}[!tb]
  \begin{minipage}[t]{.99\linewidth}
    \centering \includegraphics[width=0.8\linewidth]{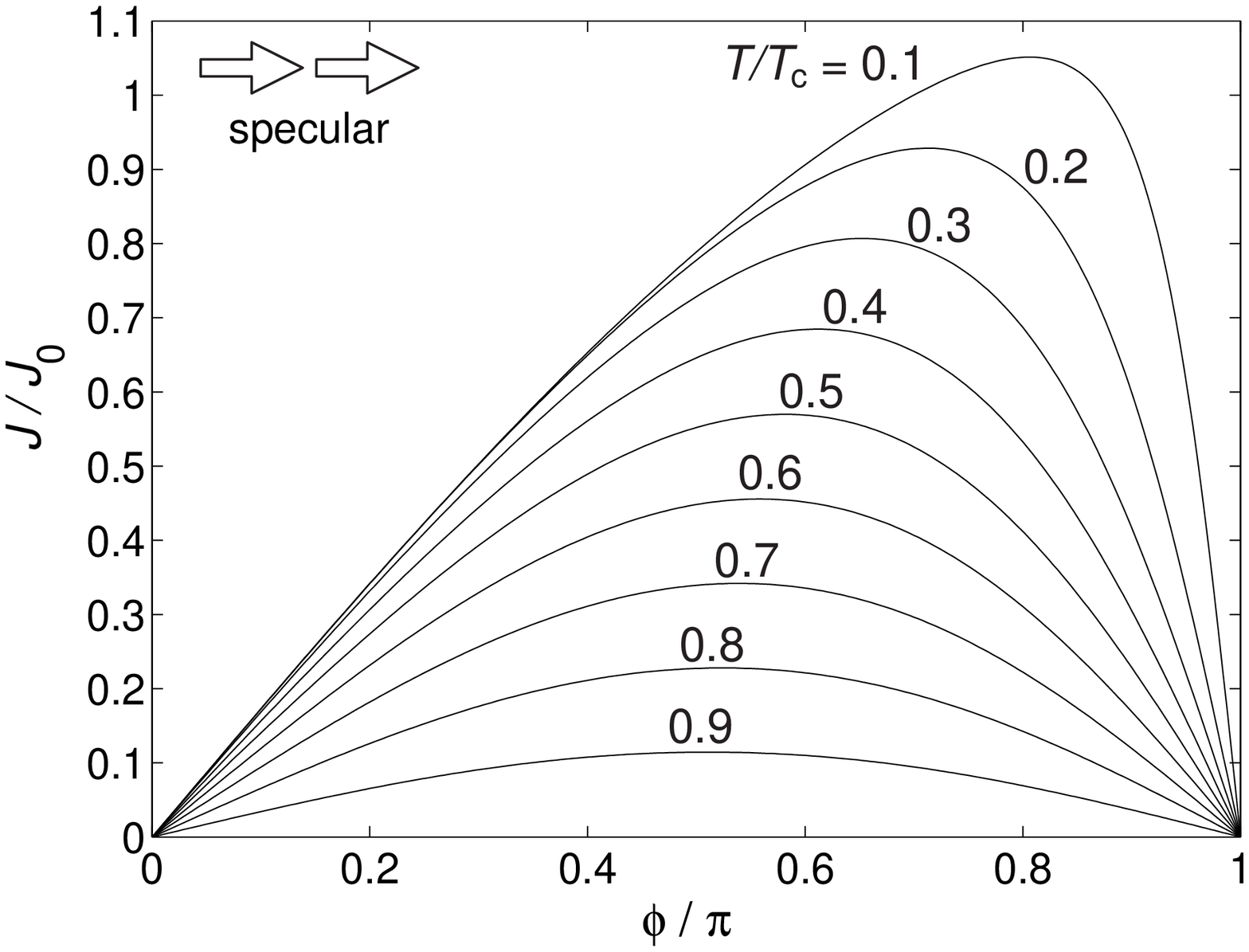}
  \end{minipage}
  \hfill
  \begin{minipage}[t]{.99\linewidth}
    \centering \includegraphics[width=0.8\linewidth]{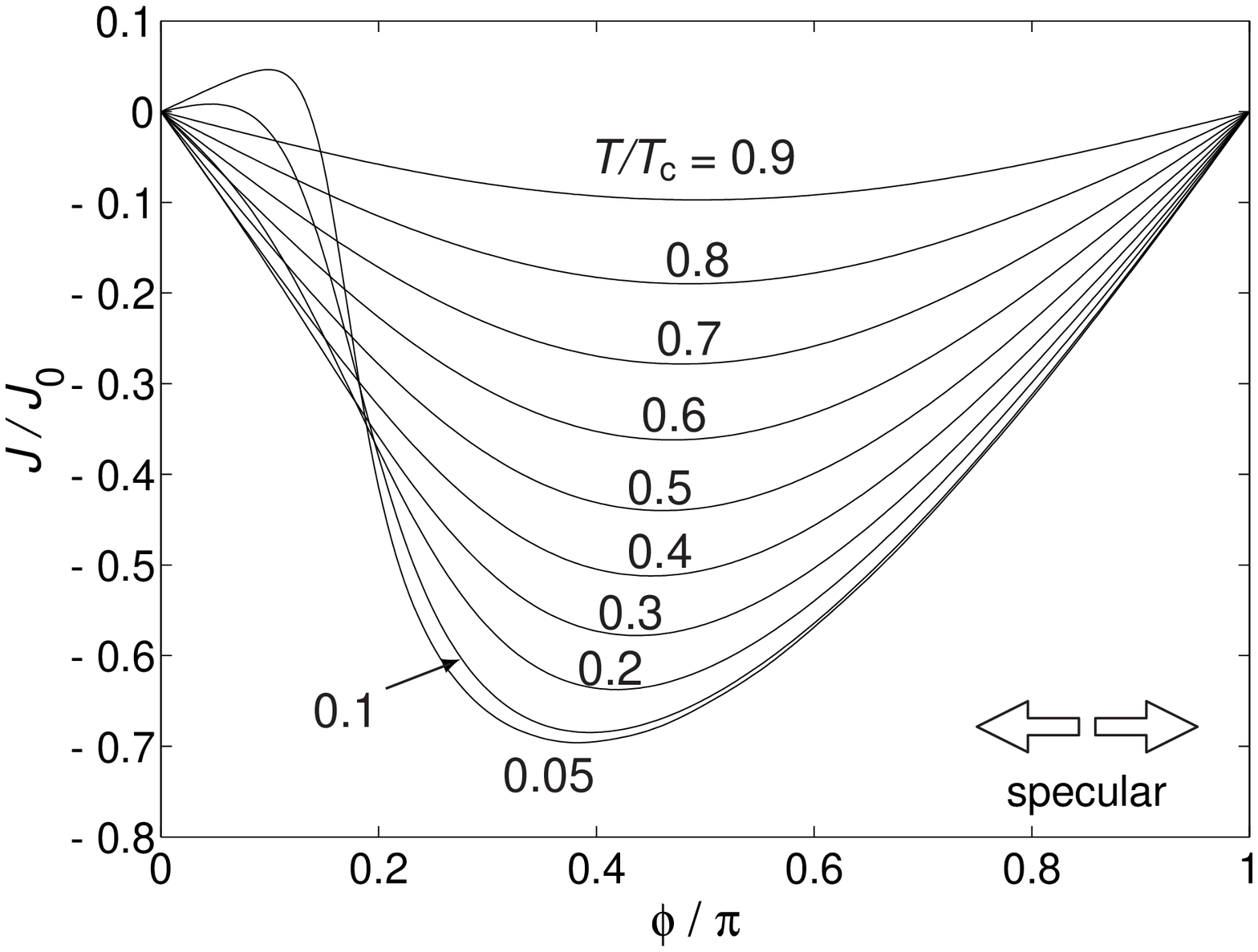}
  \end{minipage}
\caption{Isotextural current-phase relations for a pinhole in a specularly
scattering wall. The top panel corresponds to
$\nvec^L=\nvec^R$ and lower to $\nvec^L=-\nvec^R=\pm\zvec$.
$F_1^a=W/D=0$.} \label{f.siphi}
\end{figure}

\begin{figure}[!tb]
  \begin{minipage}[t]{.99\linewidth}
    \centering \includegraphics[width=0.8\linewidth]{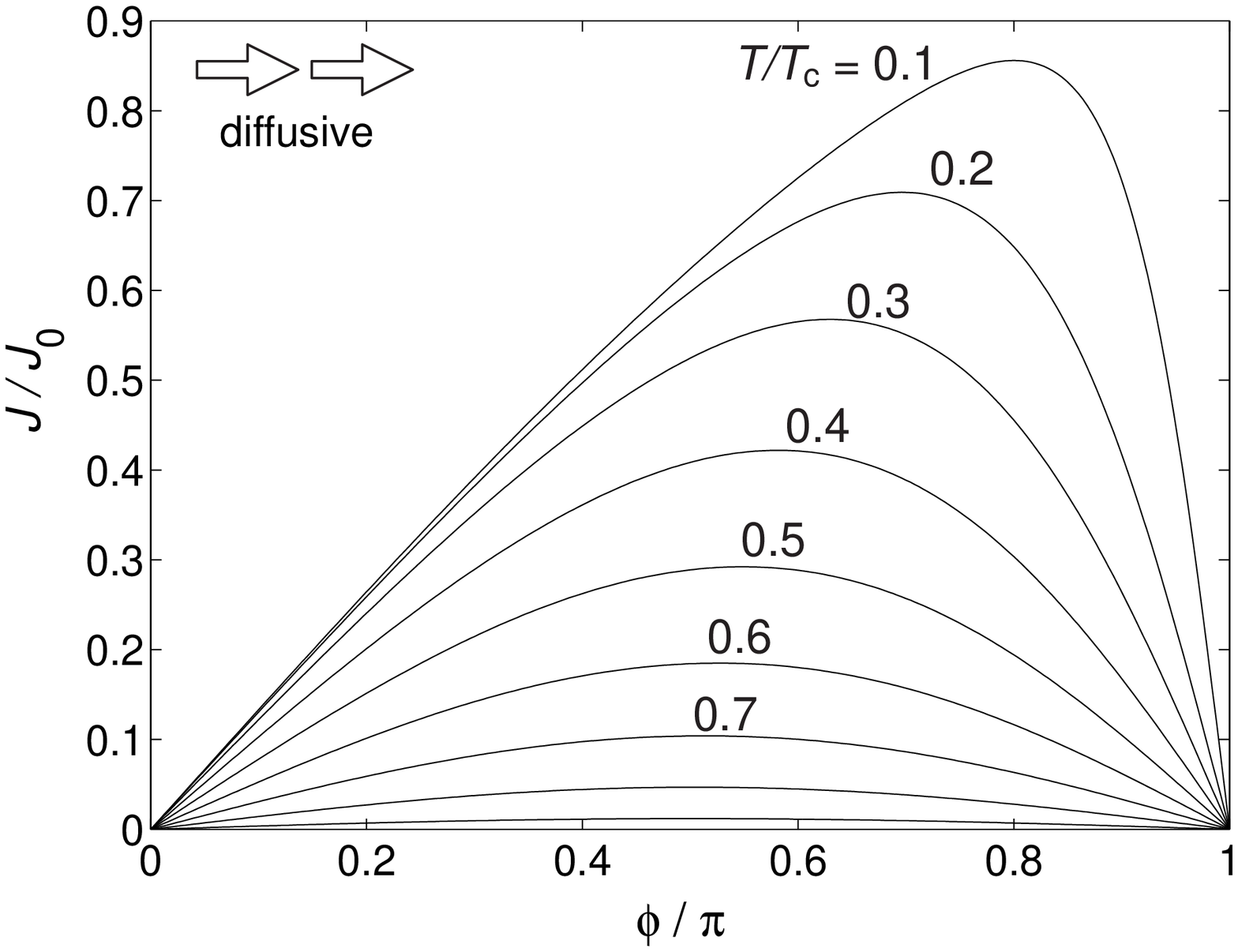}
  \end{minipage}
  \hfill
  \begin{minipage}[t]{.99\linewidth}
    \centering \includegraphics[width=0.8\linewidth]{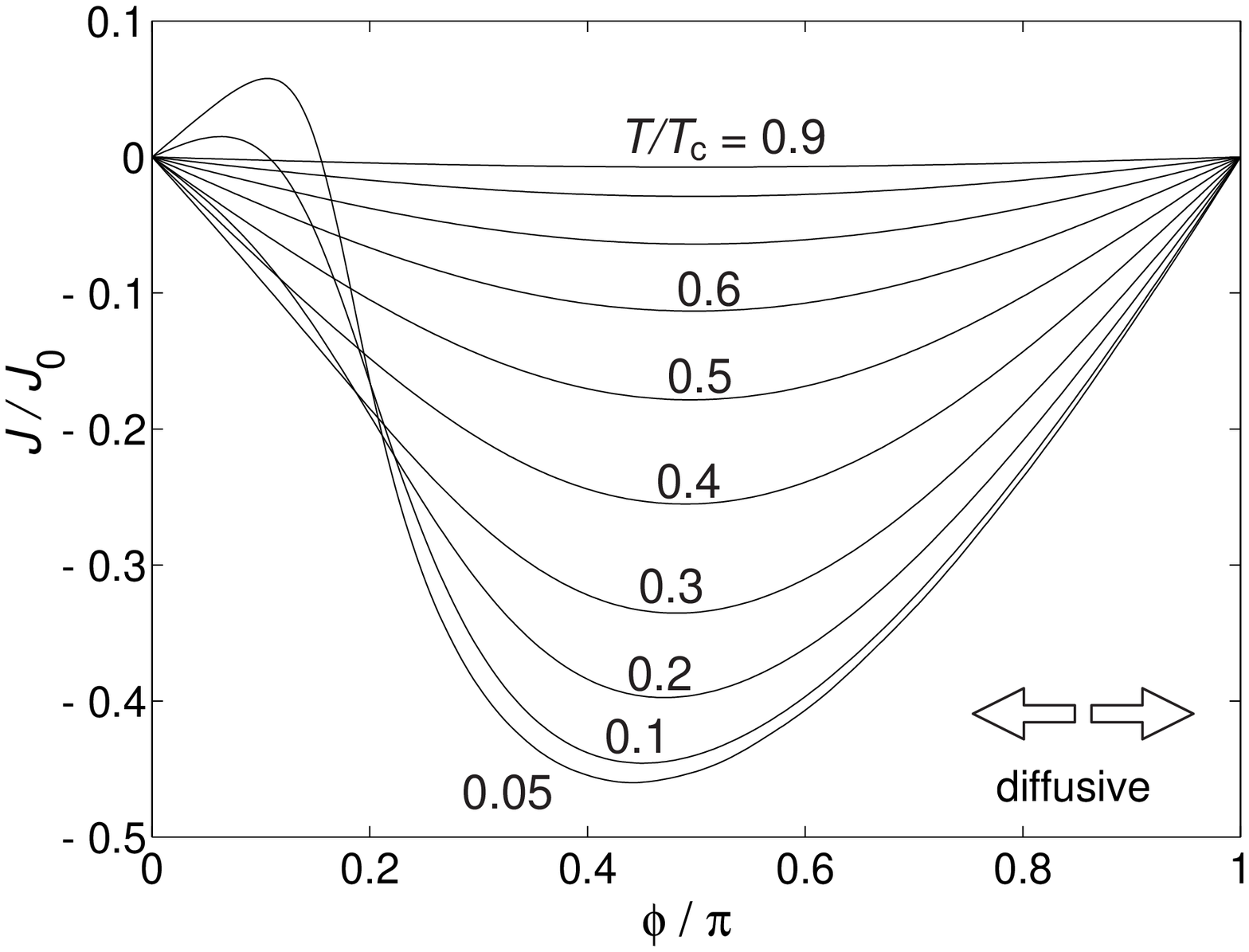}
  \end{minipage}
\caption{Isotextural current-phase relations for a pinhole in a diffusely
scattering wall. The top panel corresponds to
$\nvec^L=\nvec^R$ and lower to $\nvec^L=-\nvec^R=\pm\zvec$.
 $F_1^a=W/D=0$.} \label{f.diphi}
\end{figure}

We first study the case where
the spin-orbit rotation axes
$\nvec^{L,R}$ are perpendicular to the 
intervening thin wall. 
This situation is realistic if
the external magnetic field is small enough ($H\lesssim 1$ mT) 
and if there are no other walls with different orientations nearby. 
Four different $\nvec$ configurations are then possible, 
namely the combinations $\nvec^{L,R}=\pm\zvec$, where 
$\zvec$ is normal to the wall.  These give rise to two different
current-phase relations  corresponding to parallel ($\nvec^L=\nvec^R$)
or antiparallel ($\nvec^L=-\nvec^R=\pm\zvec$) situations. 
The parallel case is actually more general, because
the $\nvec$ vectors need not be perpendicular to the wall
to still give the same $J(\phi)$. 
The current-phase relations are shown in Figs.\ 
\ref{f.kiphi}-\ref{f.diphi} corresponding to three different surface
models. 

First we consider the case of constant order parameter.  Here
$\Deltavec(\kvec,\rvec)$ is not calculated self-consistently using any
boundary condition, but instead is assumed to have its constant bulk
form $\Delta\kvec$ all the way to the wall (no pair breaking). This is the case discussed
by Yip \cite{yippi}, and the current-phase relations shown in
Fig.\ \ref{f.kiphi} are exactly the same as obtained by him. 
The parallel case is well known: the same result was first obtained by
Kulik and Omel'yanchuk for microbridges in $s$-wave superconductors,
and it was subsequently generalized to the case of $^3$He by Kurkij\"arvi
\cite{kurkijarvi}. The new feature found by Yip is seen in the
antiparallel case: very close to $T_{\rm c}$ the $J(\phi)$ is  sinusoidal,
but at temperatures below about $0.5 T_{\rm c}$ the $\pi$ state appears: a
strong  kink and a new zero of $J(\phi)$ appear around $\phi=\pi$.

The self-consistent surface
models lead to considerable suppression of
the order parameter at the wall. As a consequence, the
$J(\phi)$'s are different for a specular surface and for a
diffusive surface, shown in Figs.\ \ref{f.siphi} and \ref{f.diphi}.
Both surface models result in qualitatively similar $J(\phi)$'s. The
parallel current-phase relations look similar as in Yip's case, although
their critical currents are slightly reduced. A clear difference is
seen  in the antiparallel cases: Firstly, the whole $J(\phi)$ appears
to be  shifted by $\pi$ so that on the phase interval $[0,\pi]$ the
current is mostly negative. Secondly, the $J(\phi)$ remains sinusoidal
down to very low temperatures.  An additional kink begins to form
only at around $0.2 T_{\rm c}$. Now the kink is also shifted from $\phi=\pi$
to $\phi=0$. We continue to call it a $\pi$ state because it
represents a local minimum of $F_{\rm J}(\phi)$ that is shifted from
the from the global minimum of $F_{\rm J}(\phi)$ by the phase difference
$\pi$.

\begin{figure}[!tb]
  \begin{minipage}[t]{.99\linewidth}
    \centering \includegraphics[width=0.9\linewidth]{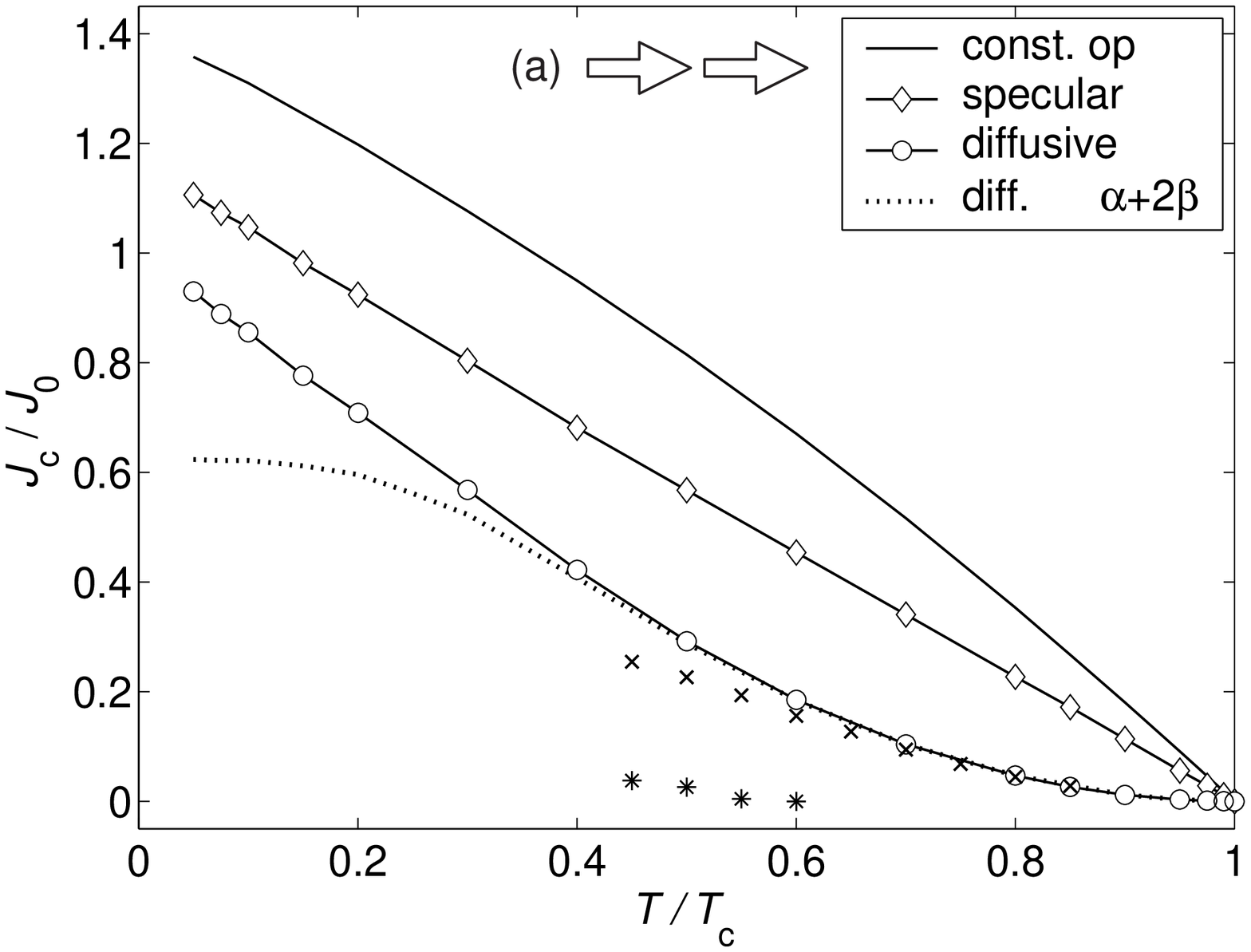}
  \end{minipage}
  \hfill
  \begin{minipage}[t]{.99\linewidth}
    \centering \includegraphics[width=0.9\linewidth]{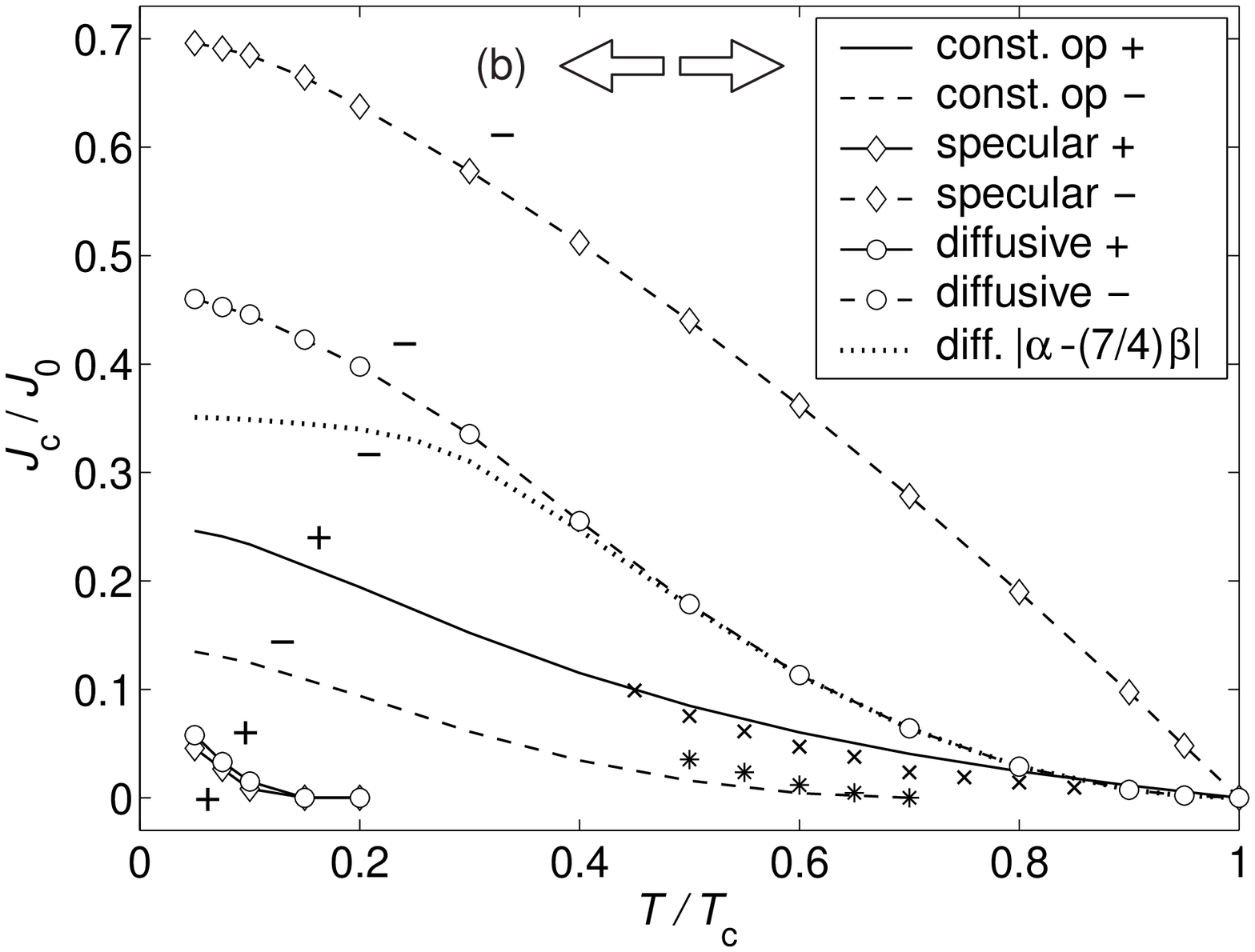}
  \end{minipage}
\caption{Critical currents as functions of the temperature. The lines
(with and without markers) are theoretical pinhole results and
separate points ($\times$, $*$) are experimental results
\protect\cite{berkeley} with $J_0=213$ ng/s (Sec.\ \ref{s.est}). For
theoretical results the upper and lower panels correspond to parallel
and antiparallel
$\nvec$ vectors, respectively. For experimental results they correspond
to the H and L states. The 
signs ($+$, $-$) denote the negative and positive extrema  of
$J(\phi)$ that appear in the case of antiparallel
$\nvec$'s. The experiment shows two extrema in both H and L cases,
whose signs are unknown.
The calculations are for $F_1^a=W/D=0$.}
\label{f.ictemps}
\end{figure}

Figure \ref{f.ictemps} shows the critical currents $J_{\rm c}$
and the possible additional extrema of $J(\phi)$ as a function of
temperature.
For parallel $\nvec$ vectors such a plot has been published in
Ref.\ \cite{tks}, but there the pinhole results were erroneous.  Close
to $T_{\rm c}$, $J_{\rm c}(T)\propto 1-T/T_{\rm c}$ for a constant order parameter and
a specular surface, and for a diffusive surface
$J_{\rm c}(T)\propto(1-T/T_{\rm c})^2$, as expected from previous calculations
\cite{kop86}.  The critical current for a constant order parameter is
always the highest and for a diffusive wall the lowest. For
antiparallel $\nvec$ vectors the roles change: the constant order
parameter case has the \emph{lowest} $J_{\rm c}$, due to the strong
cancellation between different quasiparticle directions, but the
negative extremum around $\phi=\pi$ is nearly as pronounced as the
positive one. It is  clearly visible that the other extrema appear only
at much lower temperatures for diffusive and specular surfaces.

The dotted lines correspond to the high-temperature approximations 
obtained from Eqs.\ (\ref{e.beta}) for a diffusive wall, i.e., 
$(2m_3/\hbar)(\alpha+2\beta)$ for parallel and
$(2m_3/\hbar)[\alpha-(7/4)\beta]$ for antiparallel $\nvec$'s. 
These lines follow the correct
critical currents very well down  to temperatures around
$T=0.4 T_{\rm c}$. The current-phase relations show some deviation from
the sinusoidal form at temperatures above
$0.4 T_{\rm c}$ in specular and diffusive cases, but the deviation is much
smaller than for a constant order parameter.  

\subsection{Other orientations of spin rotation axes}

\begin{figure}[!tb] 
  \begin{minipage}[t]{.99\linewidth}
    \centering \includegraphics[width=0.8\linewidth]{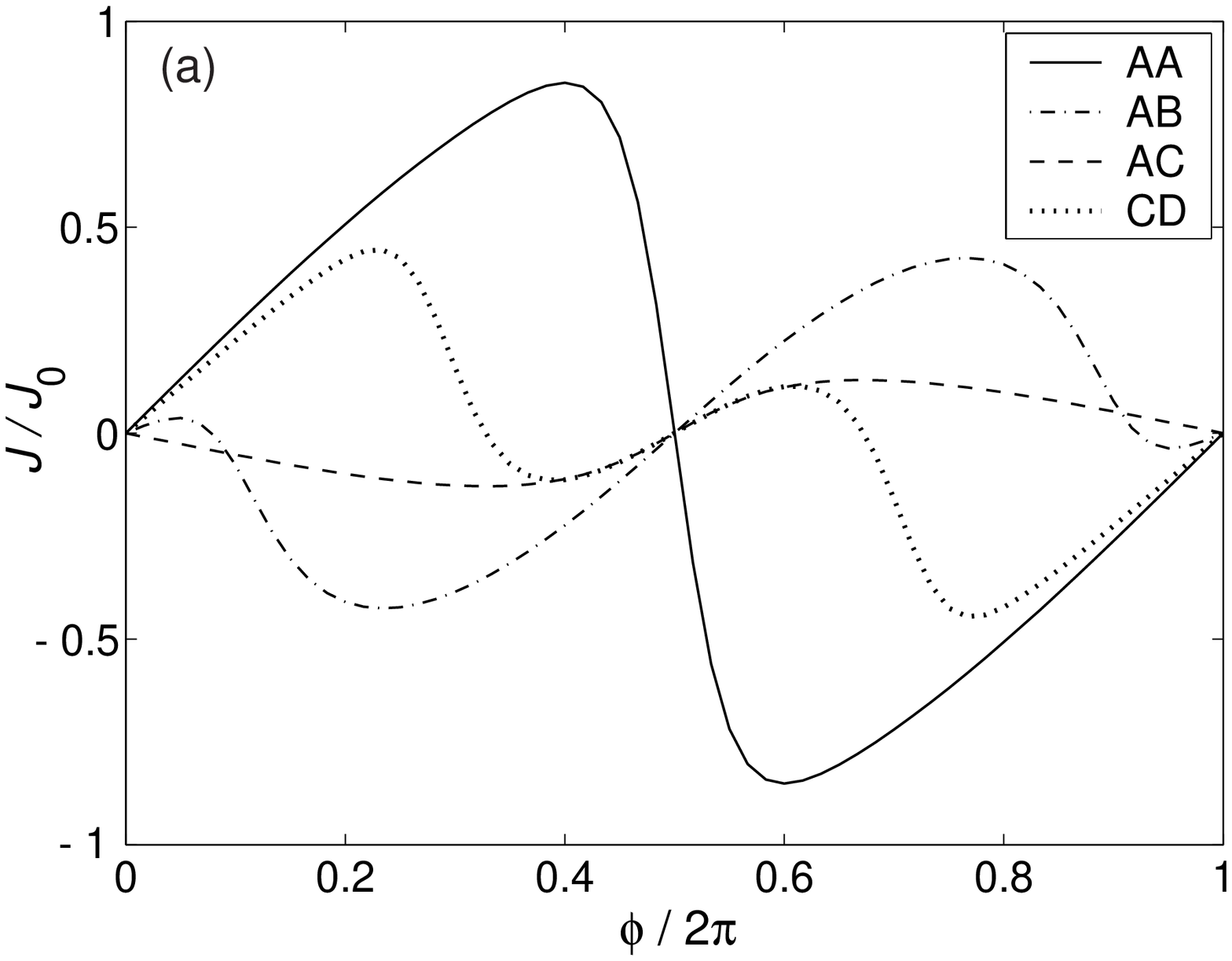}
  \end{minipage}
  \hfill
  \begin{minipage}[t]{.99\linewidth}
    \centering \includegraphics[width=0.8\linewidth]{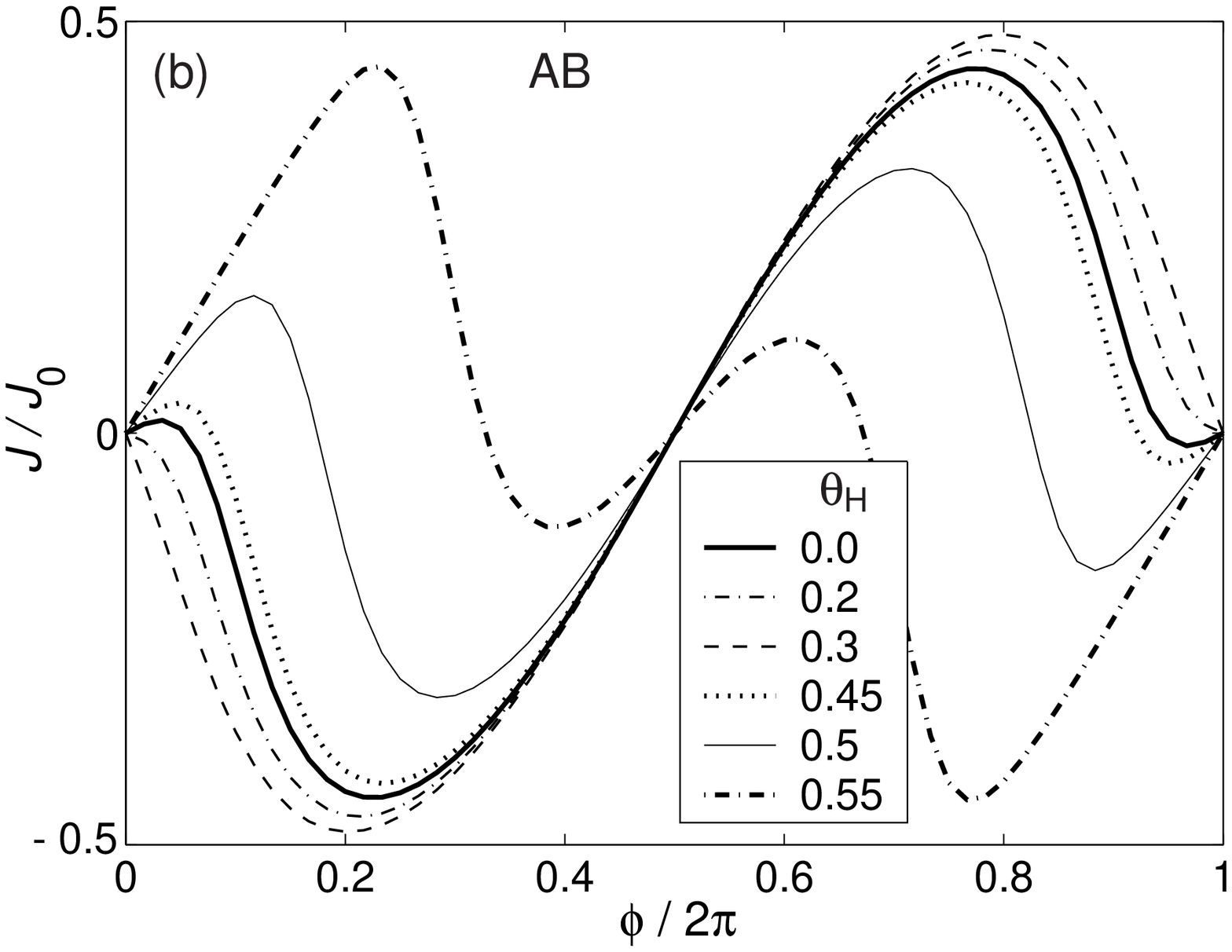}
  \end{minipage}
\caption{Examples of current-phase relations in high magnetic
field. (a) The four possible
$\nvec$ configurations which result in different current-phase
relations for magnetic field angle $\theta_H=0.45\pi$.  The
nomenclature follows the  definitions of Ref.\ {\protect \cite{yippi}}.
(b) The effect of varying $\theta_H$  in
the $AB$ configuration. The values of $\theta_H$ are given in the
legend. The curves are
for a pinhole in a diffusive  wall at $T=0.1 T_{\rm c}$ with $F_1^a=W/D=0$.}
\label{f.md01angl}
\end{figure}

In order to study other orientations of $\nvec^{L,R}$ we consider first
the case of relatively large magnetic fields, $H\gtrsim 1$ mT. The
configurations, which minimize the surface magnetic
interaction
$F_{\textrm{SH}}$ (\ref{e.sh}),  depend on the angle
$\theta_H$ between $\vecb H$ and the wall normal 
$\svec=\zvec$. Our results, obtained using the self-consistent order
parameter for a specular
and a diffusive wall, are qualitatively similar to Ref.\ \cite{yippi},
but differ in details.  Figure \ref{f.md01angl} shows the four different
$J(\phi)$'s which are possible at
$\theta_H=0.45\pi$,  corresponding to four different $\nvec^{L,R}$
configurations $AA, AB, AC$ and
$CD$, as defined in Ref.\ \cite{yippi}. Also shown is the dependence of 
$J(\phi)$ on $\theta_H$, when the $\nvec^{L,R}$ are in the $AB$ 
configuration. 
 For $\theta_H=0$ the $AB$ configuration gives
the  antiparallel case studied above. 
These should be compared with Figs. 4 and 5 of
Ref.\ \cite{yippi}, where  the same plots were given for the constant
order parameter. It can be seen that $\pi$ states are not
uncommon at low temperature $T\approx 0.1T_{\rm c}$.   

\begin{figure}[!tb]
\begin{center}
\includegraphics[width=1.0\linewidth]{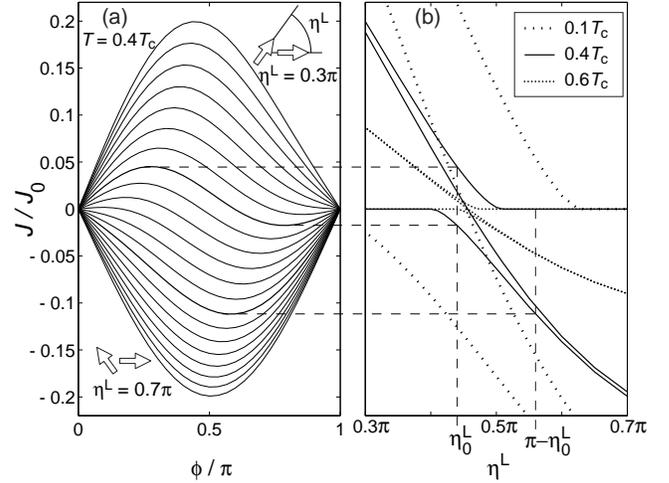}
\caption{Change of isotextural current-phase relation (a) and extremal
currents (b) as a function of $\eta^L = \arccos \hat n_z^L$ for
$\nvec^R=\zvec$. In (a) the current-phase relations are shown at
intervals
$\Delta\eta^L=0.02\pi$ at temperature $0.4T_{\rm c}$. In (b) the
extrema of $J(\phi)$
(top and bottom curves of each type) are compared with the tunneling
model results (middle curves)  at temperatures $T/T_{\rm c}=0.1$, $0.4$, and
$0.6$. The angles $\eta^L_0$ and $\pi-\eta^L_0$ correspond to 
possible bistable states, as will be discussed in Sec.\
\protect\ref{s.estiso}. The figure is calculated for diffuse
scattering and $F_1^a=W/D=0$. In the specular case the relative
deviation from the tunneling model is slightly larger ($\approx 20$
\%).}
\label{f.ictic}
\end{center}
\end{figure}

For a systematic study it seems to be more economic to specify
$\nvec^{L,R}$ directly. Let us consider the case
where the polar angle of $\nvec^L$ is changed but $\nvec^R$ is
kept constant: 
$\nvec^L\cdot\zvec=\cos\eta^L$ and $\nvec^R=\zvec$. 
It can be seen in Fig.\ \ref{f.ictic}a that $J(\phi)$ differs
essentially from the $\sin\phi$ shape around angles
$\eta^L\approx 0.46\pi$. 
The current is nearly proportional to $\sin(2\phi)$ at
$\eta^L= 0.46\pi$, and $\pi$ states are present  in the range  
$\eta^L\approx 0.42\pi\ldots 0.50\pi$ at temperature $0.4T_{\rm c}$. The
top and bottom solid lines in Fig.\ \ref{f.ictic}b show the extrema
of
$J(\phi)$ as a function of $\eta^L$. The middle solid line shows the
prediction of the tunneling model. The tunneling model (\ref{e.fj}) has
always sinusoidal isotextural $J(\phi)=J_{\rm c}\sin\phi$,
where $J_{\rm c}=(2m_3/\hbar)E_{\rm c}$ and $E_{\rm c}=\alpha
R_{\mu z}^LR_{\mu z}^R+
\beta(R_{\mu x}^LR_{\mu x}^R+R_{\mu y}^LR_{\mu y}^R)$.
Therefore, it has only a single extremum value $J_{\rm c}$ (in the
range $0<\phi<\pi$). Fig.\
\ref{f.ictic}b shows that the $\pi$ states (where two extrema appear)
take place around the configuration where $J_{\rm c}$ changes sign by
going through zero. 

Fig.\ \ref{f.ictic}b shows the extrema at three different temperatures.
At high temperatures the range where
$\pi$ states occur is very narrow. Also, in configurations showing a
$\pi$ state, both the negative and positive extrema of
$J(\phi)$ are much reduced
relative to the maximal critical currents shown in Fig.\
\ref{f.ictemps}a. Outside of the range where $E_{\rm c}\approx 0$,
$J(\phi)$ is nearly sinusoidal and the critical current is well
predicted by the tunneling model. With decreasing temperature the range
of the configurations showing a
$\pi$ state widens and the extrema get
larger. These results are not restricted to the case where
$\nvec^R=\zvec$ but are valid for general configurations of 
$\nvec^{L,R}$. 
This can be seen as a manifestation of a general  difference between
tunneling junctions and weak links, as discussed  in Ref.\
\cite{yip93}. In situations where a tunneling supercurrent  is
prohibited by symmetry, there can still be a small supercurrent flowing
in a corresponding weak link, although with some restrictions  on the
form of $J(\phi)$. 

\subsection{Nonzero $W/D$}\label{ss.wperd}

Let us consider quasiparticle scattering
inside a pinhole caused by a finite aspect ratio $W/D$. 
In a $p$-wave superfluid this can lead to a current-phase relationship
which is more complicated than some average of the ones
obtained using tunneling model and direct transmission. 
Consider a deflected trajectory (Fig.\
\ref{f.trajectory}) which is transmitted but the momentum direction is
nearly reversed. Assume $\nvec^L=\nvec^R$. If the phase difference
$\phi$ is zero, the quasiparticles of $^3$He see 
effectively a change in the sign of the order parameter
$\Deltavec(\kvec)$ along such a trajectory. However, if $\phi$ is near
$\pi$, such quasiparticles see  effectively a constant order parameter.
This means that the scattering reduces the energy at $\phi=\pi$
relative to the energy at
$\phi=0$. This could work as a mechanism for the formation of
$\pi$ states even when the spin-rotation axes on the two sides are
equal. This mechanism is closely
related to the 
$\pi$ state mechanism of Ref.\ \cite{yippi} and the effects 
discussed in Ref.\ \cite{yip95} for $sd$ contacts.

\begin{figure}[!tb]
\begin{center}
\includegraphics[width=0.8\linewidth]{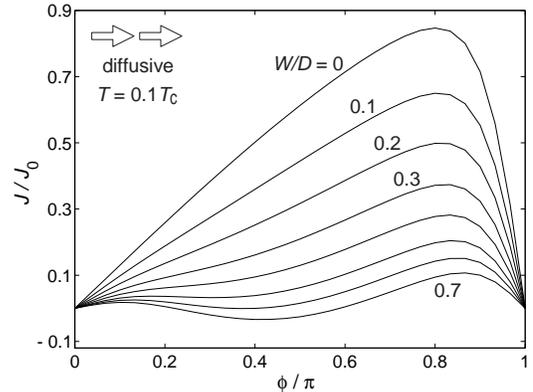}
\caption{Current-phase relations for diffusive scattering in the
aperture for $T/T_{\rm c}=0.1$, $F_1^a=0$ and $W/D=0.0\ldots0.7$ in order of 
decreasing
critical current. The $\nvec$ vectors on are parallel on the two sides
of the junction.}
\label{f.scatres}
\end{center}
\end{figure}

The effect of the scattering on the current-phase relation is shown
in Fig.\ \ref{f.scatres}.
The main change is the
decrease of the critical current. Contrary to our expectation of forming
a $\pi$ state, a dip develops at
$\phi\approx\pi/4$ with increasing $W/D$. However, this happens only
in the region of $W/D$ where the distribution (\ref{e.ddistr}) has probably
ceased to be valid. For small $W/D$ the only effect is to reduce the
critical current, and the relative initial decrease does not seem to depend
essentially on the temperature. The results
for specular scattering are very similar.

\section{Anisotextural Josephson effect} \label{s.synerg}

In the previous section we assumed that the $\nvec$ texture remains
constant when the phase difference $\phi$ is changed. In this section
we study the anisotextural Josephson effect where the texture is allowed
to change as a function of $\phi$
\cite{viljas}. We demonstrate the anisotextural effect
using a simple model. The possible existence of either
isotextural or anisotextural Josephson effects in the Berkeley
experiment \cite{bistability} is discussed in Sec.\ \ref{s.est}. 

\subsection{Phenomenological model}\label{ss.model}

We consider a planar wall separating two half-spaces of
$^3$He-B. In the absence of perturbations, the surface interaction
(\ref{e.fsd}) fixes a constant texture on both sides. The orientations
on left and right hand sides are denoted by
$\nvec^L_\infty$  and  
$\nvec^R_\infty$, respectively. They both are either parallel
or antiparallel to the normal $\zvec$ of the wall:
$\nvec^L_\infty=\pm\zvec$ and  
$\nvec^R_\infty=\pm\zvec$. Let us now place a weak link in the wall. The
Josephson coupling energy $F_{\rm J}$ (\ref{e.form2}) may now favor a
different orientation
$\nvec^{L,R}\not=\pm\zvec$ at the junction. Such a change is opposed by
a ``rigidity energy'', which consists of gradient energy and possibly
other textural energies (\ref{e.dipint})-(\ref{e.dh}). We model the
rigidity energy by a quadratic form
\begin{equation} \label{e.frigtot}
F_{\textrm{rig}}=\gamma^L(\eta^L-\eta^L_\infty)^2+
\gamma^R(\eta^R-\eta^R_\infty)^2. 
\end{equation}
Here $\eta^{L,R}$ and $\eta^{L,R}_\infty$ are the polar angles
of $\nvec^{L,R}$ and $\nvec^{L,R}_\infty$, respectively. In the
case of symmetric left and right sides, the stiffness (or rigidity) parameters
$\gamma^{L,R}$ are equal ($\gamma^L=\gamma^R=\gamma$), but the more
general form will be useful later. 

The texture is now found by minimizing the free energy 
\begin{equation}\label{e.totale} 
F=F_{\rm J}(\phi,\hat{\bf n}^L,\hat{\bf n}^R)+F_{\textrm{rig}}
(\hat{\bf n}^L,\hat{\bf n}^R, \nvec^L_\infty, \nvec^R_\infty)
\end{equation}
with respect to $\nvec^L$ and $\nvec^R$.
If the stiffness parameters
$\gamma^{L,R}$ are sufficiently small, this will lead to a $\phi$
dependent texture. This is easiest to see using the tunneling model
(\ref{e.fj}) for
$F_{\rm J}$. If $\cos\phi>0$, the minimum
$F_{\rm J}=-(\alpha+2\beta)\cos\phi$ is achieved by
$\hat{\bf n}^L=\hat{\bf n}^R$. In the opposite case $\cos\phi<0$ 
the minimum $F_{\rm J}=-(\alpha-2\beta)\cos\phi$ (assuming
$\alpha<\beta$) is achieved by $\hat{\bf n}^{L,R}=
(\mp\hat{\bf x}+\hat{\bf y}\mp\sqrt{3}\hat{\bf z})/\sqrt{5}$, for
example. Neglecting the rigidity (\ref{e.frigtot}), this leads to a
$\pi$ state with a piecewise sinusoidal current-phase relation
\begin{equation} 
J(\phi)=\cases{(2m_3/\hbar)(\alpha+2\beta)\sin\phi& for $\cos\phi>0$
\cr (2m_3/\hbar)(\alpha-2\beta)\sin\phi&for $\cos\phi<0$ \cr}. 
\label{e.idealcurrent}\end{equation}
This ideal current-phase relation is smoothed by finite
$\gamma^{L,R}$. With increasing stiffness, the texture changes less as
a function of $\phi$, and for $\gamma^{L,R}\gg\alpha$, $\beta$ the
current-phase relation reduces to the isotextural one. 

\subsection{Model parameters}\label{ss.model2}

Instead of using the tunneling model, we now assume the weak
link to consist of an array of pinholes. The phase
$\phi$ and $\nvec^{L,R}$ are assumed to be constants over the array. 
Consequently, we can use the single-pinhole results for current
(\ref{e.progcurrent}) and coupling energy  (\ref{e.final}) simply by
replacing the open area
$A_o$ by the total area of the array $A$ times $\kappa$, the fraction of
area occupied by the holes.

We start to estimate $\gamma$ including first only the gradient energy: 
\begin{eqnarray} \label{e.gene_ene}
F_{\textrm{rig}}&=&\frac{5}{8}\lambda_{\textrm{G2}}
\int\upd^3r \Big[ 16(\partial_i\hat n_j)(\partial_i\hat n_j) \nonumber\\
&&-(\sqrt{3}\nabla\cdot\nvec+\sqrt{5}\nvec\cdot\nabla\times\nvec)^2
\Big].
\end{eqnarray}
This can be obtained from the sum of gradient energies (\ref{e.sge}) and
(\ref{e.bge}) assuming
$\lambda_{\textrm{SG}}=4\lambda_{\textrm{G2}}$ and 
$\lambda_{\textrm{G1}}=2\lambda_{\textrm{G2}}$; see also
Ref.\ \cite{smith}. We shall assume that $\nvec$ varies only in one
plane, the $xz$ plane, for example. This allows one to describe $\nvec$
by its polar angle $\eta$ alone: $\nvec=\cos\eta\zvec+\sin\eta\xvec$.
In addition, we can assume $\eta$ to depend only on the radial
direction $r$ from the center of the aperture array. With these
assumptions the energy (\ref{e.gene_ene}) on one side simplifies to 
\begin{equation} \label{e.gene1}
F_{\textrm{rig}}^L=\frac{50\pi}{3}\lambda_{\textrm{G2}}\int\upd r
r^2(\partial_r\eta)^2.
\end{equation}
This is minimized by a function of the form $\eta(r)=A/r+C$, but to
avoid a divergence at $r=0$, we have to cut off the integration at some
$r$, for example at the radius of the array $R=\sqrt{A/\pi}$. As a
result one finds the form (\ref{e.frigtot}) with the stiffness
parameter 
\begin{equation} \label{e.gammapar}
\gamma=\frac{50\pi}{3}\lambda_{\textrm{G2}}R.
\end{equation}
The rigidity has also a contribution form the surface-dipole
interaction (\ref{e.fsd}). It can be
shown that this correction to
$\gamma$ is proportional to $R/\xi_{\rm SD}$ when the radius of the
array
$R$ is small compared to $\xi_{\rm SD}$ (\ref{e.xisd}). We assume this
to be the case, and therefore neglect the small correction.

In order to get numerical values for $\gamma$ (\ref{e.gammapar}) we
use the Berkeley array, where 
$R\approx0.11$ mm \cite{bistability}. The temperature dependence
is given by
$\lambda_{\textrm{G2}}$, shown in Fig.\ \ref{f.lasgp}.
The resulting $\gamma(T)$ is plotted in 
Fig.\ \ref{f.tconst}. We use this value as a standard, to which the
parameters $\gamma^{L,R}$ used in our calculations are referred to.
Note that the reference value $\gamma$ of the rigidity energy 
is larger by one order of magnitude than $\alpha$ and $\beta$, which
give the magnitude of the Josephson coupling energy (Fig.\
\ref{f.tconst}).

\begin{figure}[!tbp]
\begin{center}
\includegraphics[width=0.90\linewidth]{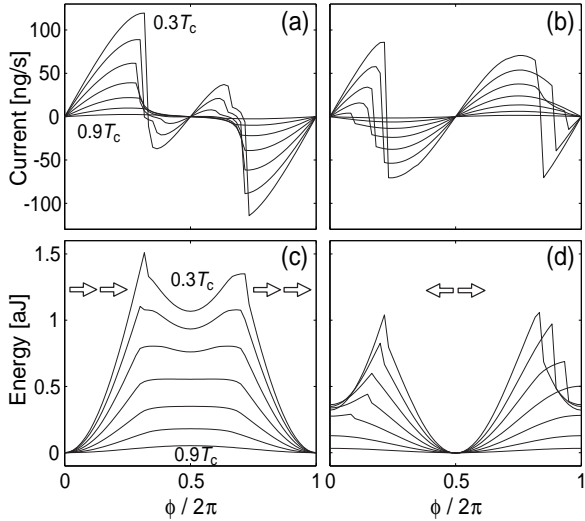}
\caption{Anisotextural current-phase $J(\phi)$ and energy-phase
$F(\phi)$  relations for bistable states $\eta^L_\infty=0$
(a,c) and
$\eta^L_\infty=\pi$ (b,d). The
curves correspond to different temperatures at intervals of 
$0.1T_{\rm c}$. The arrows indicate the
$\nvec$ orientations of in the zero branch, where
$\nvec^L=\nvec^L_\infty$ and $\nvec^R=\nvec^R_\infty$. The rigidities on
both sides are equal,
$\gamma^L=\gamma^R=0.1\gamma$ and $\eta^R_\infty=0$.  The
weak link parameters and
$\gamma$ (Fig.\ \protect\ref{f.tconst}) are evaluated for the Berkeley
array  (Sec.\ \protect\ref{s.est}). The wall is assumed diffusive,
$F_1^a=W/D=0$. }
\label{f.wdratold}
\end{center}
\end{figure}

In order to calculate the current, the free energy (\ref{e.totale}) 
should be minimized with respect to three angles parametrizing
$\nvec^{L,R}$:
$\eta^L,\eta^R$ and the relative azimuthal angle $\chi$. 
The results of such a minimization
for 
$\gamma^L=\gamma^R=0.1\gamma$ are shown in Figure \ref{f.wdratold}.
In creating these curves, we proceeded from left to right, using the
minimum angles ($\eta^L, \eta^R,\chi$) of each $\phi$ step as the 
initial guess for the next step. The panels on
the left are for the parallel case ($\eta^L_\infty=\eta^R_\infty=0$).
For
$\phi$ from
$0$ to approximately $\pi/2$, the vectors $\nvec^{L,R}$ remain exactly 
perpendicular to
the wall. At low temperatures, a discontinuous 
jump to another branch of $J(\phi)$ occurs at around 
$\phi=\pi/2$, where $\nvec^{L,R}$
are tilted form their original perpendicular positions. With increasing
$\phi$ there is a jump back from this $\pi$ branch so that for
$\phi>3\pi/2$ the minimum solution again corresponds to perpendicular
$\nvec$'s. The panels on the right are for the antiparallel case
($\eta^L_\infty=\pi$, $\eta^R_\infty=0$). Here the $\pi$ branch, where
$\nvec$'s are tilted, occurs not at 
$\phi\approx\pi$ but at $\phi\approx 0$. The fact that the curves are
not (anti)symmetric with respect to $\phi=\pi$ indicates that the jumps
between different branches are hysteretic. 

We have studied the effect of increasing $\gamma^L=\gamma^R$
from the value $0.1\gamma$. We find that the $\pi$ state first
disappears in the antiparallel case and then also in
the parallel case. For example, the $\pi$ state in the
parallel case [defined as positive
$J'(\pi)$] disappears when $\gamma^{L,R}\approx 0.2\gamma$ at
$T=0.4T_{\rm c}$. The anisotextural effect on the current-phase relation
still continues up to $\gamma^{L,R}\approx 0.7\gamma$. 
Since the coupling (\ref{e.final}) scales with
$A_o=A\kappa=\pi R^2
\kappa$ and stiffness (\ref{e.gammapar}) with $R$, we get the following
necessary condition for the appearance of $\pi$ states in the parallel case at
temperatures $T \gtrsim 0.4T_{\rm c}$
\begin{equation} \label{e.synecond} R\kappa> 0.5\ \mu\textrm{m}.
\end{equation} Thus, the larger the radius $R$ of the array and
the higher the ratio of the open area $\kappa$, the better are the
chances of realizing the anisotextural $\pi$ state.
At higher temperatures it will be more difficult, due to the 
relative strength 

It is interesting to compare the condition (\ref{e.synecond}) with the
Ginzburg-Landau calculation in a single large hole 
\cite{viljas}. There $\pi$ states could be seen for hole radii 
$R> 5.5\xi_{\textrm{GL}}$. Setting $\kappa=1$ and extrapolating
$\xi_{\textrm{GL}}(0.4T_{\rm c})= 8.8$ nm \cite{hydro}, we have the
condition
$R \kappa > 0.48\ \mu$m in surprising agreement with the pinhole
result  above.

\subsection{Asymmetric case}\label{ss.asummetric}

\begin{figure}[!tbp]
\begin{center}
\includegraphics[width=0.95\linewidth]{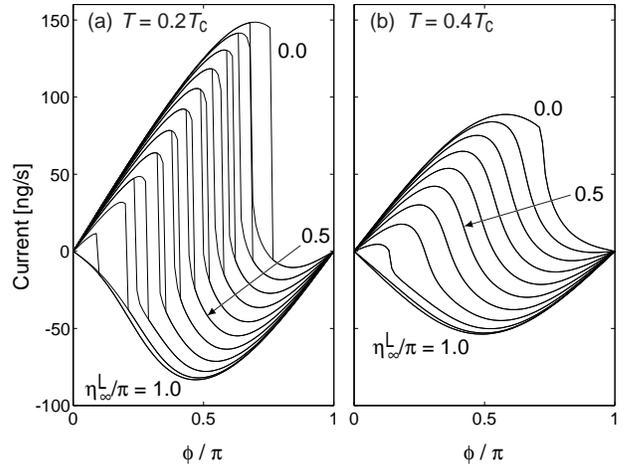}
\caption{Anisotextural $J(\phi)$ for different asymptotic angles
$\eta^L_\infty$ at temperatures $0.2T_{\rm c}$ (a) and $0.4T_{\rm c}$
(b). The other parameters are
$\eta^R_\infty=0$,
$\gamma^L=0.3\gamma$, and $\gamma^R=\gamma$. The different curves
correspond to $\eta^L_\infty$ at intervals of $0.1\pi$.
Other parameters are the same as in Fig.\ \protect\ref{f.wdratold}. }
\label{f.asymtemps}
\end{center}
\end{figure}

We go slightly beyond the simple model of an infinite planar
wall introduced in Sec.\ \ref{ss.model}. Firstly, we allow the
asymptotic directions $\nvec^{L,R}_\infty$ to be arbitrary. Secondly, 
the stiffness coefficients $\gamma^{L,R}$ can be different.
In Fig.\ \ref{f.asymtemps} we have studied different values of
$\eta^L_\infty$ while $\eta^R_\infty=0$,
$\gamma^L=0.3\gamma$, and $\gamma^R=\gamma$. 
Here the extreme curves represent parallel and antiparallel
$\nvec^{L,R}_\infty$. It can be seen that neither of them have a $\pi$
state at
$0.4T_{\rm c}$ and only the parallel case has one at $0.2T_{\rm c}$.
However, at intermediate $\eta^L_\infty$ the $\pi$ states still persist
in a wide range of $\eta^L_\infty$. The current-phase relationships  are
not hysteretic at high temperature, but hysteresis develops at lower
temperatures. 
Figure \ref{f.asymtemps} should be compared with the
corresponding isotextural Fig.\
\ref{f.ictic}. In the isotextural case the $\pi$ states occur only in
limited range, and there is no hysteresis.

\begin{figure}[!tbp]
\begin{center}
\includegraphics[width=0.90\linewidth]{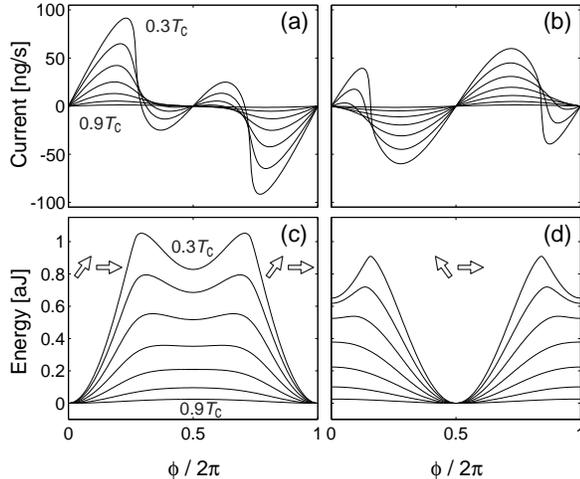}
\caption{Anisotextural $J(\phi)$ and 
$F(\phi)$ for bistable states $\eta^L_\infty=0.3\pi$ (a,c) and
$\eta^L_\infty=0.7\pi$ (b,d). The textures have different rigidities
$\gamma^L=0.3\gamma$, $\gamma^R=\gamma$ and $\eta^R_\infty=0$. 
Other parameters are the same as in Fig.\ \protect\ref{f.wdratold}. }
\label{f.wdrat}
\end{center}
\end{figure}

The states with $\eta^L_\infty=\pi/2\pm{\rm constant}$ are
expected to be degenerate in the absence of the Josephson coupling
($\eta^R_\infty=0$). The currents and energies for one pair of
such bistable states are shown in Fig.\ \ref{f.wdrat}.
The important difference to Fig.\ \ref{f.wdratold} is that the curves
are smooth and there is no hysteresis. In the case of parallel and
antiparallel
$\nvec_\infty$'s the symmetry is spontaneously broken in the $\pi$
branch, whereas the tilted $\nvec_\infty^L$ already breaks the symmetry,
and thus the $\pi$ state can develop continuously.  

\subsection{Discussion}

The results of Figs.\ \ref{f.wdratold}-\ref{f.wdrat} contain both
the isotextural and anisotextural mechanisms of $\pi$ states.
However, practically the same results can be obtained all the way down
to $T=0.4T_{\rm c}$ by using the tunneling model $F_{\rm J}$
(\ref{e.fj}), which excludes the isotextural $\pi$ state. As discussed
above, the tunneling model fails at high temperatures only if the
Josephson energy is close to zero. The
minimization procedure seems to avoid
such a situation, and thus the tunneling
model gives a good description of the anisotextural pinhole array at
temperatures above $T\approx0.4T_{\rm c}$. 

Here we comment our earlier calculations of the anisotextural Josephson
effect in Ref.\ \cite{viljas}. Firstly, there the tunneling model was
used instead of the general pinhole result. Secondly, the scattering
within the hole was partially taken into account by reducing the
transmission by factor $p(\vartheta)$ (\ref{e.probability}), but the
deflected trajectories were completely neglected. Thirdly, the 
coupling parameters $\alpha$ and $\beta$ were scaled by different
factors, which is not possible when using the general pinhole
result. Taken together, these explain why Fig.\ \ref{f.wdratold} is
different from Fig.\ 2 in Ref.\ \cite{viljas}.

To be accurate, we should add to the preceding 
calculation the effect of the flow in the macroscopic region far away
from the weak link. Above the phase difference $\phi$ is defined between
the macroscopic-mesoscopic borders of the two sides (Fig.\
\ref{f.junctionregions}). The ``true'' phase difference between 
the infinities can be obtained by adding 
$(2m_3/\hbar)J/(\pi R\rho_s)$ to $\phi$, where the correction
assumes a radial flow outside of the radius $R$.
Correspondingly the total 
energy gets an additional contribution $J^2/(2\pi
R\rho_s)$. These rescalings do not affect the results qualitatively.

The anisotextural model, as described above, assumes a vanishingly small
external magnetic field. Qualitatively, it is easy to see what the
effect of a strong magnetic field would be. In a field $H\gg1$ mT, the
strongest interaction affecting the texture is the surface-magnetic
term $F_{\rm SH}$ (\ref{e.sh}). If the coupling energy scale is much
smaller than this, then the texture will be fixed to some minimum of
$F_{\rm SH}$ and will not depend on the phase difference. A strong
magnetic field therefore suppresses any anisotextural $\pi$
state and only the isotextural mechanism remains. 

In order to estimate the critical field, we equate the change in
magnetic surface energy $\Delta F_{\rm SH} \approx dA{H}^2$ at the
junction with the gain of energy $\Delta E$ between the $\pi$ state
and ``0 branch'' at
$\phi=\pi$. Thus $H_{\rm c}\approx \sqrt{\Delta E/(dA)}$. For the Berkeley
array
\cite{bistability} $\Delta E$ can be estimated from the energy-phase
graphs of Figs.\
\ref{f.wdratold} and
\ref{f.wdrat} or from the experiments.  At $T\approx0.45 T_{\rm c}$ the
values are on the order of 
$\Delta E\approx 0.1\ldots0.5$ aJ, which yields
the order of magnitude
$H_{\rm c}\approx 10\ldots 50$ mT.

\section{Analysis of the Berkeley experiment} \label{s.est}

\begin{figure}[!tbp]
\begin{center}
\includegraphics[width=0.90\linewidth]{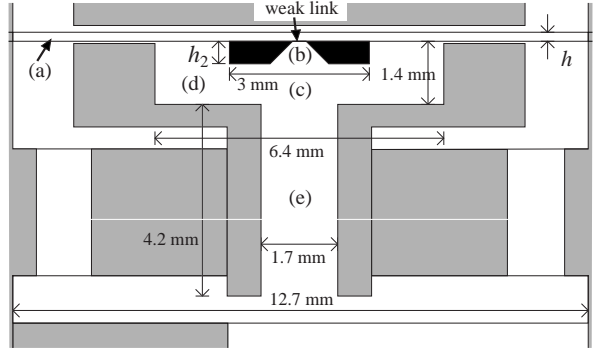}
\caption{Part of the experimental cell in Ref.\
\protect\cite{bistability}. Most of the structure has rotation
symmetry around the vertical axis. The ``pillbox'' (a) between two
flexible membranes (diameter
1.27 cm, distance $h=0.14$ mm) forms a volume that is
connected to the rest of $^3$He volume only through the weak link.
The weak link is made in a silicon chip attached to the lower membrane. 
The chip is 
$h_2= 0.5$ mm thick and has a square window opening from 250 $\mu$m
$\times$ 250
$\mu$m to some 0.95 mm $\times$ 0.95 mm at the lower chip
surface. The figure is based on a drawing supplied by S.
Pereverzev. }
\label{f.pillest}
\end{center}
\end{figure}

We now turn to an analysis of the Berkeley experiment 
\cite{bistability}. There the weak
link consists of a square array of $65\times 65$ holes. They were etched
in a $50$ nm thick silicon chip with a hole spacing of 3
$\mu$m, making the area of the array 195 $\mu$m $\times$
195 $\mu$m. The holes were nominally squares  $100$ nm $\times$ $100$
nm. However, flow resistance measurements in the normal state seem to
indicate a somewhat larger apertures $115$ nm $\times$ $115$ nm
\cite{normalstate}, and these larger values are used in all numerical
estimates in this paper.  A sketch of the experimental cell  is shown in
Fig.\ \ref{f.pillest}.
The scattering properties of $^3$He quasiparticles
from the silicon chip are not known, but most surfaces are generally
believed to be diffusively scattering.  The magnetic field is believed
to be small,
$H\ll 1$ mT, and the pressure is 0 bar. We further assume the system to
be in thermal equilibrium. 

The central experimental findings are the bistability and the
existence of $\pi$ states
\cite{bistability}. Bistability means that the system can randomly
choose between two alternative states, characterized by high (H) and low (L)
critical currents.  Both of these states show $\pi$ states. The measured
extremal currents are plotted in Fig.\ \ref{f.ictemps}. 

There are two major difficulties in applying the theory presented in
this paper to the Berkeley experiment. Firstly, since $\xi_0=77$
nm, the holes are too large to be
pinholes. Also, the holes are too small for the Ginzburg-Landau
calculations to be reliable \cite{thu88,viljas}. We use the pinhole
model because more accurate calculations would be much more
demanding.  Moreover, due to the approximate
nature of our pinhole calculations for finite aspect ratio
$W/D$, we will only use the pinhole theory in the limit
$W/D=0$, although experimentally
$W/D\approx 0.4$. Another reason for using $W/D=0$ is that
the measured critical currents are clearly larger than
calculated for pinholes with $W/D=0.4$ (Sec.\ \ref{ss.wperd}).
The second major difficulty is that the cell is complicated, and its
dimensions are on the same order of magnitude as
$\xi_{\textrm{SD}}$ (\ref{e.xisd}). Instead of a proper calculation of
the texture, we will make some simple estimates and introduce one
adjustable parameter. 

\subsection{Isotextural Josephson effect}\label{s.estiso}

Here we consider the case where the Josephson coupling can be
considered as a weak perturbation, which does not affect the texture
on either side of the weak link. For small magnetic field, the
dominant orienting effect on $\nvec$ comes from the surface-dipole 
energy (\ref{e.fsd}). In region (a) of Fig.\ \ref{f.pillest} this
clearly favors the uniform texture
$\nvec=\pm\hat{\bf z}$, where $z$ is along the axis of the cell. The
situation is more complicated on the other side of the junction. The
axial orientation is preferred in the wide cylindrical region (d). The
narrow cylindrical region (e) favors 
$\nvec\perp\zvec$. The tendencies from regions (d) and (e) compete in
region (c) below the chip, and affect the texture in the window region
(b). Let us assume that the minimization of textural energies
(\ref{e.dipint})-(\ref{e.dh}) favors at the junction a particular
orientation of
$\nvec^L=\nvec^L_0$ with polar angle $\eta^L_0$. Then there must be a
degenerate textural state with
$\nvec^L=-\nvec^L_0$, which corresponds to the polar angle
$\eta^L=\pi-\eta^L_0$. There can be additional degeneracy with respect
to the azimuthal angle. 

When we add the pinhole Josephson coupling, the degenerate states above
give rise to precisely two different current-phase relationships if
$\eta^L_0\not=\pi/2$. These two are obtained by studying
configurations with 
$\eta^L=\eta^L_0$ and
$\eta^L=\pi-\eta^L_0$ with a fixed $\nvec^R=\hat{\bf z}$.
Configurations of this kind were studied above in Fig.\ \ref{f.ictic}. 
They can in principle explain both the bistability
and the existence of
$\pi$ states.
Quantitative comparison with experiment gives, however, very poor
agreement. The problem is to fit the four experimental points in Fig.\
\ref{f.ictemps} at any temperature with the four extrema in Fig.\
\ref{f.ictic}b at angles $\eta^L_0$ and
$\pi-\eta^L_0$ using $\eta^L_0$
as a fitting parameter. One such construction at $T=0.4T_{\rm c}$ is
shown by vertical lines in Fig.\ \ref{f.ictic}b. In that case, one
finds the L state with a
$\pi$ state, but then no $\pi$ state appears in the H state, and all
currents are more than by a factor of 2 too small. Alternatively, if one
tries to fit the critical current in the H state, one has to approach
the parallel and antiparallel states, where no $\pi$ states appear at
this temperature (Figs.\
\ref{f.siphi}-\ref{f.ictemps}).

\subsection{Anisotextural Josephson effect}

For the anisotextural Josephson effect we have to calculate the
rigidity energy (\ref{e.frigtot}). The region (a) is much thinner than
$\xi_{\rm SD}$ and therefore we have to consider a two-dimensional
texture instead of the 3D in Sec.\ \ref{ss.model2}. In the present case
the surface interaction is important. We find that $\gamma^R$ is
proportional to $\lambda_{\rm G2}h/\ln(\sqrt{h\xi_{\rm SD}}/R)$
instead of $\lambda_{\rm G2}R$ (\ref{e.gammapar}),  but these two happen to
be of the same order of magnitude. Thus the texture in region (a) is
rather stiff, and we assume $\gamma^R\approx\gamma$. This makes the
texture essentially fixed.

In order to analyze the other side (L), we first consider a conical
region between two radii, $R_1$ and $R_2$. Otherwise, we use the same
approximations as for the half space in Sec.\
\ref{ss.model2}. We get the rigidity energy 
\begin{equation} \label{e.cone}
F_{\rm rig}^{(1-2)}=
\frac{50\pi}{3}\lambda_{\textrm{G2}}(1-\cos\theta)
\frac{R_1R_2}{R_2-R_1}[\eta(R_1)-\eta(R_2)]^2,
\end{equation}
where $\theta$ is the opening angle of the cone.
This reduces to the previous result (\ref{e.gammapar}) in the limit 
$\theta=\pi/2$ and $R_2=\infty$.
We now consider the regions (b) and
(c) as three consecutive conical regions. The middle one corresponds
to the conical region of the chip with $\theta=\theta_w\approx
38^\circ$, the two others have $\theta=\pi/2$. We neglect the effect of
regions (d) and (e) and assume that region (c) extends to infinity. For
the combined stiffness of these regions we get the estimate   
\begin{equation} \label{e.estim2}
\gamma^L\approx
\frac{50\pi}{3}\lambda_{\textrm{G2}}\left[\frac{1}{R}+
\frac{\cos\theta_w}{1-\cos\theta_w}\left(
\frac{1}{R_1}-\frac{1}{R_2}\right)\right]^{-1},
\end{equation}
where $R_1$ and $R_2$ are the radii of the inner and outer edges of
the conical region measured from the weak link.
Substituting the numerical values we find 
$\gamma^L\approx 0.31\gamma$, where $\gamma$ is the half-space value
(\ref{e.gammapar}). Thus this side is considerably softer
than the other. 

The calculations for the present parameters have already been done
in Sec.\ \ref{ss.asummetric}. The results are presented in 
Fig.\ \ref{f.asymtemps} for different angles $\eta^L_\infty$. A
representative pair of bistable states is presented in Fig.\
\ref{f.wdrat}. The calculated current-phase relations are very similar
to those found experimentally
\cite{bistability}. The $\pi$
states are present in both bistable states. The critical current in the
H state (identified with $\eta^L_\infty\approx 0.3\pi$) is very close.
Some 
differences can be resolved. For example, the $\pi$ state is too strong
in the theoretical H state, and the critical current in the theoretical
L state is slightly too large. The only fitting parameter here
is the textural angle $\eta^L_\infty$, and it can be seen from
Fig.\ \ref{f.asymtemps} that $\eta^L_\infty=(0.5\pm 0.2)\pi$
roughly represents the best overall fit to the
experiments \cite{bistability}. Taking into account that the
experimental apertures are not pinholes and that only one fitting
parameter is involved, the agreement between the
anisotextural theory and the experiment is amazingly good.

The anisotextural Josephson model gives several predictions that can be
tested experimentally. The shape of the current-phase relationship
crucially depends on the number of parallel apertures and also on their
spacing. It depends on the geometry of the cell surrounding the weak
link. It depends on the magnetic field. The current-phase relationships
become hysteretic at low temperatures. All
these dependences are either absent or very different in the
isotextural model. The dependences can be
quantitatively extracted from the theory presented above.
 
None of the predictions have been studied experimentally yet, 
excluding possibly the hysteresis. The first paper \cite{berkeley} reported
discontinuous jump to the
$\pi$ branch at low temperature $0.28 T_{\rm c}$. The later paper
\cite{bistability} reported continuous current-phase relationships
but only at higher temperatures $>0.45 T_{\rm c}$. However, there was
a change in the experimental set up between these observations, which
may have affected the results. The hysteresis should also show up 
as additional dissipation, but no detailed theory
yet exists \cite{dissipation}.

\section{Conclusions} \label{s.conc}

We have presented above a fairly complete study of the dc Josephson
effect in
$^3$He-B using the pinhole model. We have derived a general energy
functional for the pinhole coupling energy. A computer program has been
constructed to calculate the coupling energy and currents in a general
case. We have plotted the isotextural current-phase relationships for
various cases. Besides the mass current there is also a spin-current,
but that has not been examined in this work. In contrast to most pinhole
calculations, we consider also a finite aspect ratio of the hole, but no
extensive studies have been made because of the approximate nature
of the model. We have calculated some surface
parameters of $^3$He-B.

We have studied the anisotextural Josephson effect. The previous
tunneling model calculations are generalized to arrays of pinholes.
The general conditions for the anisotextural effect
are studied. It is found that the anisotextural Josephson effect
depends sensitively on parameters like the dimensions and the number of
holes, the surrounding geometry, and the magnetic field. 

The theory is applied to explain the experimental observations made at
Berkeley. We compare the experiment with both iso- and anisotextural
models. Both mechanisms can in principle explain the bistability
and
the $\pi$ states. In quantitative comparisons there is one
adjustable parameter describing the texture. Quantitative comparison
with the isotextural model gives poor agreement, but a very good
agreement is found with the anisotextural model. New experiments
should be made to confirm the identification of the anisotextural
Josephson effect.

Experiments on the Josephson effect in $^3$He-B have also been done using
a single aperture \cite{singleaperture}. Both $\pi$ states and
multistability are observed. The pinhole theory can hardly be applied
to this case because the aperture is much larger than the coherence
length $\xi_0$. Also, the distinction between iso- and anisotextural
effects is not well defined for a single aperture. The Ginzburg-Landau
calculations should be more accurate here, but unfortunately they have
been done systematically only for parallel $\nvec$'s \cite{viljas}.

\begin{acknowledgments}
The authors would like to thank O. Avenel, J. C. Davis, A. Marchenkov, 
R. Packard, S. Pereverzev, R. Simmonds, and E. Varoquaux for discussions. We thank
the Berkeley group for sending experimental data prior to their publication. The
Center for Scientific Computing is acknowledged for providing 
computer resources for the numerical calculations during this project.

\end{acknowledgments}

\end{document}